\documentclass[12pt]{article}

\usepackage{gensymb}
\usepackage{url}
\usepackage{amssymb,amsmath, amsfonts}
\usepackage{cases}
\usepackage{latexsym}
\usepackage{graphicx}
\usepackage{relsize}   
\usepackage{natbib}
\usepackage[includeheadfoot,left=1.in,right=1.in,top=1.in,bottom=1.in,bindingoffset=0.in,nohead]{geometry}
\usepackage[onehalfspacing]{setspace}  
\usepackage{booktabs}
\usepackage{appendix}  
\usepackage[pdfborder={0 0 0}]{hyperref}  
\usepackage{float}
\usepackage{xcolor} 
\usepackage{caption}    
\usepackage{subcaption} 
\usepackage{minibox}

\usepackage{multirow}
\usepackage{makecell} % Line-break in Table-Cell

\usepackage{colortbl}										% Make new colors
\definecolor{france}{rgb}{0.19,0.55,0.91}			% 
\definecolor{sienna}{rgb}{0.91,0.45,0.32} 			% 
\usepackage{fancyhdr}									% Page number in top-right corner
\fancypagestyle{pagenumbertopright}{
	\fancyhf{}
	\fancyhead[R]{\thepage}
	
}
\usepackage{pdflscape}									%	Make a landscape page

\usepackage[shortlabels]{enumitem}

\begin{document}

\title{When Will Arctic Sea Ice Disappear?\\ Projections of Area, Extent, Thickness, and Volume}

\smallskip

	\author{Francis X. Diebold\\University of Pennsylvania \and Glenn D. Rudebusch \\Brookings Institution \and Maximilian G\"obel\\University of Lisbon \and Philippe Goulet Coulombe\\University of Quebec at Montreal \and Boyuan Zhang\\University of Pennsylvania \\$~$}
	
	\maketitle

	\begin{spacing}{1}
		
		\noindent \textbf{Abstract}: Rapidly diminishing Arctic summer sea ice is a strong signal of the pace of global climate change. We provide point, interval, and density forecasts for four measures of Arctic sea ice:  area, extent, thickness, and volume. Importantly, we enforce the joint constraint that these measures must \emph{simultaneously} arrive at an ice-free Arctic. We apply this constrained joint forecast procedure to models relating sea ice to atmospheric  carbon dioxide concentration and models relating sea ice directly to time. The resulting ``carbon-trend" and ``time-trend" projections are mutually consistent and predict a nearly ice-free summer Arctic Ocean by the mid-2030s with an 80\% probability. Moreover, the carbon-trend projections show that global adoption of a lower carbon path would likely delay the arrival of a seasonally ice-free Arctic by only a few years.  
		
		\thispagestyle{empty}
				
		\bigskip
		
	\bigskip

	\footnotesize
				
		\noindent {\bf Acknowledgments}:   For helpful comments we thank, without implicating,  the Editor, Co-Editor, and two anonymous referees, as well as Eric Hillebrand, Walt Meier, Aaron Mora, Felix Pretis, Richard Startz, and Zack Miller, as well as participants at seminars and conferences.  For research assistance we thank Jack Mueller and Gladys Teng.

		\bigskip

		\noindent {\bf Key words}: Climate change, cryosphere, climate prediction, climate forecasting, carbon dioxide concentration, carbon emissions
		
				\bigskip

		{\noindent  {\bf JEL codes}: Q54, C51, C52, C53}

		\normalsize
		
	\end{spacing}
	
%	\clearpage
%	
%	\thispagestyle{empty} 
%	\setcounter{tocdepth}{3}
%	\tableofcontents
%	
	
 \clearpage
	\setcounter{page}{1}
	\thispagestyle{empty}

\section{Introduction}  \label{intro}

Understanding and forecasting  the climatic observational record (COR) is critical for a variety of tasks -- notably, the planning of future mitigation and adaptation infrastructure and investment.
The COR has many diverse elements, such as temperature (means, extremes, variability, etc.), extreme events (storms, floods, droughts, wildfires, etc.), and other physical attributes (sea level, cloud coverage, glacial extent, etc.).  These elements differ importantly at a spatial level, and the Arctic region has emerged as a salient focal point for investigations of ongoing climate change.  In particular, Arctic melting is a conspicuous effect of climate change, which has  warmed the Arctic two to three times faster than the global average.  The melting Arctic also promotes additional future climate change, 
in part due to an ``albedo'' feedback amplification loop as more reflective Arctic sea ice is replaced by darker open water. Along with global effects, a melting Arctic also has pervasive regional consequences. For example, in the Arctic region, less sea ice will reduce the cost of shipping, promote the extraction of natural resources, and expand tourism. Together, at a global and regional level, a melting Arctic will result in widespread and transformative impacts, opportunities, and risks \citep{Alvarez2020, LynchNorchiLi2022, BrockMiller2023}.

Large-scale climate models have been the source of much of the forward-looking global and Arctic climate analysis. These models attempt to capture the fundamental physical drivers of the earth's climate with much spatial and temporal detail. Such structural models are invaluable for understanding climate variation; however, from a forecasting perspective, climate models have been less adept and, in particular, have largely underestimated the amount of lost sea ice in recent decades (e.g., \citealp{StroeveEtAl2012}; \citealp{Rosenblum2017}). 
As a practical complementary approach, \emph{statistical} projections from parsimonious dynamic time series representations often produce forecasts that are at least as accurate as detailed structural models (\citealp{DRice}).  COR forecasting can be facilitated by using reduced-form econometric/statistical  methods, which  are designed to  summarize the relevant data patterns (correlations) regardless of the deep underlying causal climatic mechanisms. \label{enthusiasm} This reduced-form agnosticism is valuable for COR forecasting because the granular climate dynamics are incompletely understood -- witness the fifty or so different CMIP6 climate models and their widely-differing projections (e.g., \citealp{NotzSIMIP2020}). Specifically, for forecasting Arctic sea ice, there is already some evidence that small-scale econometric/statistical models can have some success (\citealp{Wangetal2016, DRice}).

\label{coverage}   In this paper, we therefore attempt to push forward the econometric/statistical analysis of the long-run future evolution of Arctic sea ice. Of particular interest are probability assessments of the timing of an ice-free Arctic, an outcome with vital economic and climate consequences (\citealp{Jahn2016}). A key contribution of our analysis is that we take a multivariate approach to the various measures of Arctic sea ice. Specifically, we consider three aspects of pan-Arctic sea ice: surface coverage (measured in two ways -- area and extent, to be defined shortly), thickness, and volume. In theory, these three measures of sea ice should be perfectly related.  Volume would equal the product of surface coverage and thickness, so modeling any two would imply a specification for the third.  In practice, however, each sea-ice indicator is based on a blend of  pure observations and model interpretations and interpolations. As a result, measurement differences and errors introduce discrepancies between the various indicators and drive wedges between them, even if the amount of measurement error has generally diminished over time as observational and modeling methods have improved. Thus, the various measures of sea ice are not perfectly related, and it may be beneficial to consider all of them.

We  characterize the various measures of September Arctic sea ice as time series processes and examine the likely future path toward seasonal ice-free conditions with continued global warming. September is the month with the least seasonal ice, and we focus on projecting the timing of the first ice-free September, exploring both a literally ice-free Arctic (IFA) and an effectively or nearly ice-free Arctic (NIFA).\footnote{We follow the usual convention and define NIFA as sea-ice coverage of less than $10^6\, {\rm km}^2$ as described below.}  
   We model sea-ice area,  extent, thickness, and volume, and we implement multivariate models that constrain these measures to reach zero simultaneously. We anchor our analysis on area, sequentially considering pairwise blends of area with each of the other indicators. The area-extent pair is essentially based on only observed data; the area-thickness pair integrates the two key indicators  characterizing sea ice;  and the area-volume pair is closest to an all-indicator analysis that incorporates area, thickness, and extent.

We implement our  constrained joint forecasting procedure in ``carbon-trend" models that relate sea ice to atmospheric carbon dioxide (CO$_2$) concentration (with this time series denoted as \textit{CO2C}). This allows us to condition our forecasts on various carbon scenarios.  We also check our carbon-trend results against similarly-constrained  ``time-trend" models that  relate sea ice simply to time.  Finally, we verify the consistency of the two approaches, and we use them to make probabilistic forecasts of the arrival years of NIFA and IFA.

There is of course a large  literature on  measurement and modeling of declining Arctic sea-ice extent and area, and insightful overviews include   \cite{StroeveEtAl2012}, \cite{SeaIceBookCh4}, and  \cite{IPCC2021_Ch9}. For analyses of the measurement and modeling of declining thickness and volume, see, e.g., \cite{StroeveNotz2018} and \cite{SeaIceBookCh5}.  We will  have more to say throughout the paper about the literature on Arctic sea ice and how our results relate to it. For now, however, we emphasize again that our multivariate modeling of area, extent, thickness, and volume -- respecting in particular the fact that if and when they vanish, they must vanish simultaneously -- sets our analysis apart. 
\label{differences} In particular, the present paper is quite different from our earlier research, \cite{DRice}, which compares structural climate models to reduced-form statistical models based on polynomial trends.  The present paper avoids any statistical vs structural forecast model comparisons. Instead, it dives more deeply into the statistical analysis and modeling, incorporating much more information to form statistical forecasts: (1) more sea-ice indicators, (2) bivariate models with equal-IFA constraints imposed, and  (3) conditioning on a key geophysical covariate (CO$_2$ concentration), instead of just a generic polynomial trend on a single sea-ice indicator. 

We proceed as follows. In section \ref{data}, we introduce our four sea-ice indicators.  In section \ref{carbonpro},  we model and forecast the  indicators  using regressions on  atmospheric  CO$_2$ concentration (``carbon trends"),  producing point and interval forecasts.  In section  \ref{trendpro},  we provide a major robustness check by modeling and forecasting  using direct time trend regressions   (``time trends"), assessing the consistency of the  carbon-trend and time-trend  forecasting approaches.  In section \ref{Disappear}, we use both approaches to provide full probabilistic forecasts of the two objects of ultimate interest: the arrival years of IFA and NIFA.  In section \ref{emissions}, we provide a second major robustness check by redoing most of the carbon-trend analysis using an alternative carbon measure, cumulative anthropogenic  CO$_2$ emissions (\textit{CO2E}), assessing the consistency of the two carbon-trend versions.   We conclude in section  \ref{conclsec}.

\section{Measures of Arctic Sea Ice}  \label{data}

We consider four key aspects of Arctic sea ice: surface coverage (of which there are two measures, area and extent), thickness, and volume. Our analysis focuses on the seasonal minima for these indicators, specifically, the September monthly average at a pan-Arctic scale.\footnote{See Appendix \ref{data_app} for data sources and details.}

\subsection{Surface Coverage (Area and Extent)}

Arctic sea-ice surface coverage has been well measured using satellite-based passive microwave sensing since late 1978.  The Arctic's surface is divided into a grid of cells, and for each cell, satellite sensors measure the amount of reflected solar radiation.  This reflected energy varies from cell to cell depending on the relative cell coverage of sea ice and open water, as the latter absorbs more radiation. The resulting measured ``brightness'' is converted into a fractional measure of sea-ice ``concentration'' for cell $i$ at time $t$, denoted $c_{i, t}$.  The individual $c_{i, t}$ are then aggregated into two standard, somewhat different measures of overall sea-ice coverage.

The first of these, ``sea-ice area" ($SIA$), sets measured concentration in each cell to zero if it is below .15 and retains the satellite-recorded value otherwise: 
\begin{equation}
  \label{SIAdef}
c_{i, t}^{SIA} = \left \{
\begin{array}{c}
0, ~{\rm if}~c_{i, t} \le .15\\
c_{i, t}, ~ {\rm otherwise}   . 
\end{array}
\right.
\end{equation}
All grid cell areas are then multiplied by $c_{i, t}^{SIA}$, and their weighted sum over the entire Arctic is denoted $SIA_t$. As a result, $SIA$ is the measured coverage of all Arctic grid cells with at least 15 percent sea-ice concentration.

The second measure of coverage, ``sea-ice extent" ($SIE$), also sets measured concentration below .15 to zero but otherwise adjusts it upward to full coverage, or 1: 
\begin{equation}
  \label{SIEdef}
c_{i, t}^{SIE} = \left \{
\begin{array}{c}
0, ~{\rm if}~c_{i, t} \le .15\\
1, ~ {\rm otherwise}   . 
\end{array}
\right.
\end{equation}
In this case, all grid cell areas are multiplied by $c_{i, t}^{SIE}$, and their weighted sum is total Arctic $SIE_t$. Thus, $SIE$ is the total area of all of the ocean grid cells that are measured to have at least 15\% sea-ice coverage.  

The  up-rounding in the construction of $SIE$ serves as a bias correction to offset satellite sensors' tendency to mistake shallow pools of sea-ice surface melt for open sea.\footnote{See \url{https://www.ncdc.noaa.gov/monitoring-references/dyk/arctic-ice} for details.}  There are costs and benefits of such a bias correction, and both $SIA$ and $SIE$ are widely employed in research and forecasting. We will examine both, using September monthly average $SIA$ and $SIE$ data from 1979 to 2021 from the National Snow and Ice Data Center (NSIDC).

The up-rounding in the construction of $SIE$ also implies that by definition $SIE \ge SIA$, and indeed, in the historical data, $SIE$ has been substantially higher than $SIA$. For our analysis, the crucial definitional constraint is that these two measures of sea-ice coverage can only equal zero under the same condition, that is, when \emph{all} Arctic grid cells have no more than fifteen percent coverage ($c_{i, t}\le .15$ for all \emph{i}). In that case, the two measures will be equivalent, so $SIA=SIE=0$.  This constraint that both indicators must reach zero at the same time will be the crucial timing restriction that allows us to construct a joint projection of  pan-Arctic September $SIA$ and $SIE$.

\subsection{Thickness}

Our next sea-ice indicator is thickness, denoted $SIT$.  Sea-ice thickness -- the distance between the ocean underneath and any snow coverage or air above the ice -- is  challenging to measure but allows a fuller understanding of sea-ice conditions.\footnote{See, for example, \cite{Bunzeletal2018}, \cite{Chevallieretal2017}, \cite{SeaIceBookCh5}, \cite{NotzSIMIP2020}, and \cite{Zygmuntowskaetal2014}.} 
In the past, data on sea-ice thickness have been obtained via a variety of methods including submarine records, ocean floor buoys, bore holes, helicopter and sled surveys, and satellite measures, but these sources have varying accuracy and availability by time and location. A popular alternative to these disparate thickness measures is produced by the Pan-Arctic Ice-Ocean Modeling and Assimilation System (PIOMAS)  from the  University of Washington's Polar Science Center  \citep{ZhangRoth03}.  
The PIOMAS thickness indicator incorporates (or assimilates) data such as the NSIDC sea-ice concentration measure as well as atmospheric information such as wind speed and direction, surface air temperature, and cloud cover. However, the PIOMAS data are importantly driven by an ocean and sea-ice physical model that characterizes how sea-ice thickness evolves over location and time in response to dynamic and thermodynamic forcing and mechanical redistribution such as ridging.
That is, the PIOMAS thickness data are effectively a  model construct, although they have been largely validated against the available observational data.  For example, \cite{Selyuzhenoketal2020}, building on \cite{Schweigeretal2011}, report that the spatial pattern of PIOMAS ice thickness agrees well with observations derived from in situ and satellite readings, and \cite{Labeetal2018} note that decadal trends in thickness appear realistically reproduced. Surveys that reach a similar conclusion include \citet[section 8.4]{SeaIceBookCh8} and \citet[section 5.6]{SeaIceBookCh5}.

\subsection{Volume}

Our final sea-ice indicator is volume, $SIV$, which is also supplied via the model-based PIOMAS measurement. Specifically, across a grid of cells, this series combines the PIOMAS thickness and the sea-ice coverage data to produce sea-ice volume cell by cell. Of course, in theory, volume, which represents the entire mass of sea ice, should be the most comprehensive indicator of Arctic sea ice.  On the other hand,  $SIV$, which is produced as the product of surface coverage and thickness, inherits measurement error from both of those series and appears to be subject to the most uncertainty in measurement.

\subsection{Remarks on a joint zero-ice constraint}

As noted above, our joint estimation approach relies on a simultaneous zero-ice constraint on more than one measure of sea ice.  The two sea-ice coverage series, $SIA$ and $SIE$, share data sources and definitions that effectively hard-wire or force such a simultaneous ice-free Arctic. A similar exact relationship is less assured for thickness and volume as these depend on model-based imputations.  Furthermore, $SIT$ and $SIV$ do not impose the 15\% cutoff used in $SIA$ and $SIE$, which potentially could lead to later ice-free dates for thickness and volume (though our empirical results below suggest the opposite timing). However, given that \emph{all} four indicators depend on the amount of sea-ice coverage, it is likely that $SIA$, $SIE$, $SIT$, and $SIV$ will all reach zero very closely in time if not simultaneously, and we will assess the imposition of that constraint in the estimation of sea-ice dynamics.

\section{Linear Carbon-Trend Models, Fits, and Forecasts} \label{carbonpro}

Here we explore  the empirical relationships between the four Arctic sea-ice indicators and  atmospheric CO$_2$ concentration.  We then use these representations to forecast the likely progression of a melting Arctic and, specifically, the years of first IFA and NIFA.\footnote{\label{greenhouse}Although other greenhouse gases, such as methane, are also responsible for global warming, CO$_2$ is by far the most important. Moreover, exploration of a broad aggregation of greenhouse gases that takes into account the varying global warming impacts and duration of different gases -- so-called ``CO$_2$ equivalent" --  produced  results similar to those reported below.}

\subsection{Atmospheric CO$_2$ Concentration Data}

\begin{figure}[t]
	\caption{Atmospheric CO$_{2}$ Concentration Scenarios}
	\begin{center}
		\includegraphics[trim={0mm 70mm 0mm 60mm},clip,scale=.2]{{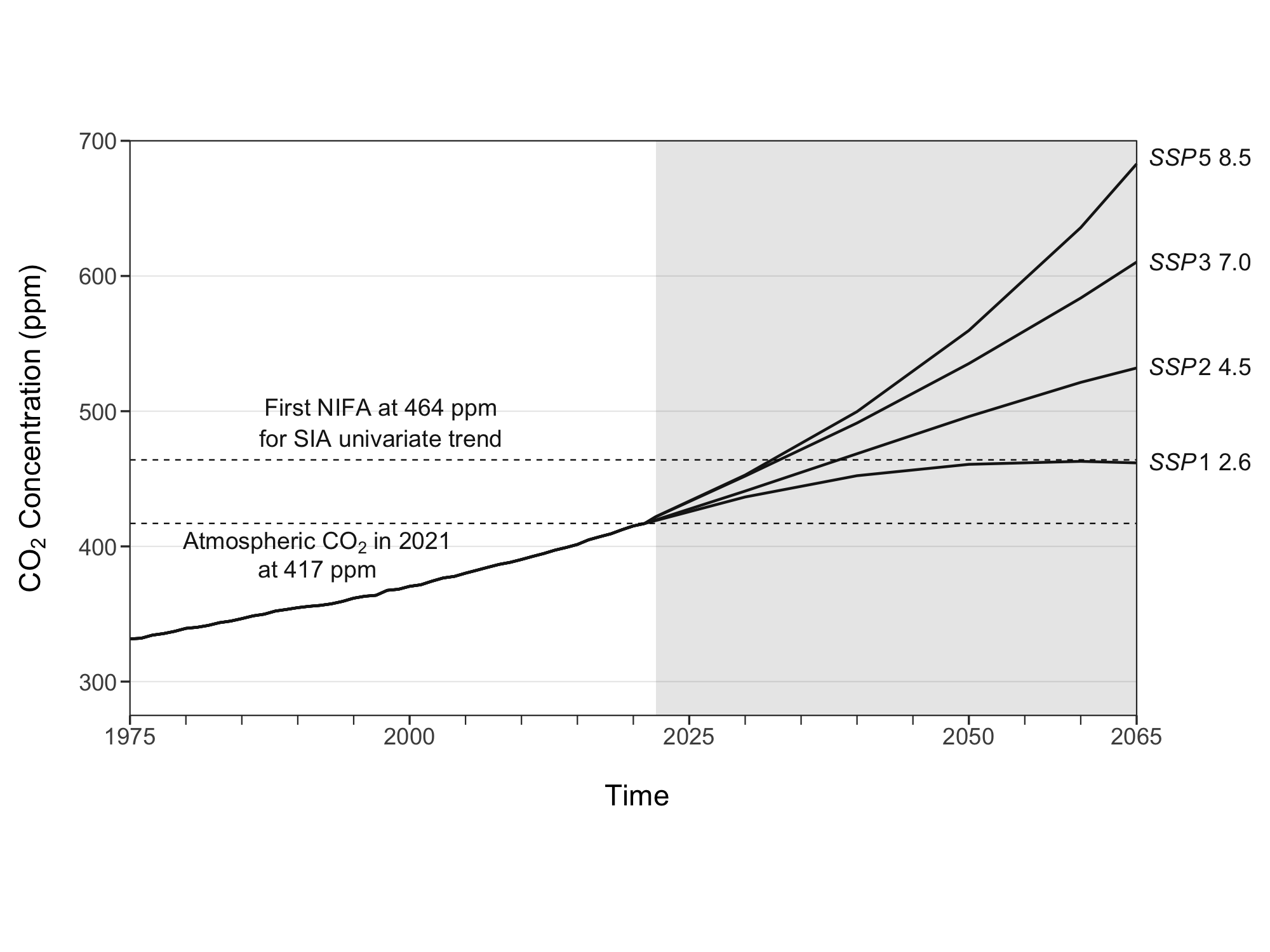}}
	\end{center}
	\label{fig:emissions}
	\begin{spacing}{1.0}  \noindent \footnotesize Notes: We show historical data and four projected scenarios for atmospheric CO$_{2}$ concentration, measured in parts per million (ppm). The historical period (to 2021) is unshaded, and the out-of-sample period is shaded.  The lower dashed horizontal line denotes the current concentration level, and the upper dashed horizontal line provides our estimate of the CO$_{2}$ concentration level associated with first NIFA using the $SIA$ univariate carbon-trend specification.  See text for details.
	\end{spacing}
\end{figure}

\label{concentration} We measure CO$_2$ using readings on atmospheric concentration rather than cumulative emissions, which is a widely used alternative (e.g., \citealp{NS2016}). Atmospheric CO$_2$ concentration has notable advantages in this regard. It is a direct measure of the amount of heat-trapping gasses in the atmosphere -- the source of the melting Arctic. In particular, unlike cumulative emissions, atmospheric concentration can account for time-varying CO$_2$ absorption rates, so as the efficacy of natural terrestrial and oceanic carbon sinks change over time, concentration can better capture the extent to which greenhouse gases are driving climate change. In addition, atmospheric CO$_2$ concentration, which is measured by direct air sampling, is subject to little measurement error. By contrast, fossil CO$_2$ emissions are indirectly calculated from energy, fuel use, and cement production data, while emissions from land-use changes are based on deforestation estimates.  

Our atmospheric CO$_{2}$ concentration series (measured in parts per million, ppm) is shown in Figure \ref{fig:emissions}.  The historical data (1979-2021) are from the NOAA Global Monitoring Laboratory, as measured at the Mauna Loa Observatory in Hawaii. The four projected scenarios (2022-2100, shaded) are from the SSP Public Database.\footnote{\label{sspmap}See Appendix \ref{data_app} for details.}  These are the standard SSP scenarios denoted {SSP}1 2.6 (which has very low emissions or radiative forcing), {SSP}2 4.5 (low emissions), {SSP}3 7.0 (medium), and {SSP}5 8.5 (high).\footnote{See, for example, \cite{IPCC2021tech}.} They are distinguished by the amount of progress made in reducing further concentration increases, as shown by the variation in their upward curvature.  We focus on the {SSP}3 7.0 scenario as a baseline but also show that our results are qualitatively robust to use of the low and high emissions scenarios.

\subsection{The Carbon-Trend Model}  \label{carbtrend}

Various researchers -- including  \cite{Johannessen2008}, \cite{NS2016}, and \cite{StroeveNotz2018} -- have identified a linear empirical relationship between observed Arctic sea-ice area and CO$_2$.\footnote{Some researchers measure CO$_2$ as atmospheric concentration and others as cumulative emissions.  As discussed above, we employ the former; however, as shown in section  \ref{emissions}, our results are robust to use of cumulative emissions.}  This linear carbon-trend relationship, which fits remarkably well in the observed data, can be expressed as:
\begin{equation}
  \label{equ:co2_uni}
	  SIA_t = a +  b ~ \text{\textit{CO2C}}_{t} + \varepsilon_t  ,
\end{equation}
where $SIA_t$ is sea-ice area, \textit{CO2C$_{t}$} is atmospheric CO$_2$ concentration, and $\varepsilon_t$ represents deviations from the linear fit.

 The regression intercept, $a$, calibrates the level of sea-ice coverage, and the slope, $b$, provides a measure of the Arctic-carbon response. A negative value for $b$ captures the diminishing coverage of Arctic sea ice as greenhouse gases accumulate in the atmosphere.\footnote{\label{SSW}As regards possible endogeneity of $CO2C$, note that we are not attempting to estimate a structural coefficient.  Rather, we are simply estimating a reduced-form best linear predictor under quadratic loss, for which least squares in levels is always consistent, regardless of the endogeneity/exogeneity status of the regressors, and regardless of whether the series are stationary, deterministically trending, stochastically trending (integrated), or cointegrated \citep{SSW}.}  Equation (\ref{equ:co2_uni}) also can be used to characterize the occurrence of first IFA in terms of \textit{CO2C}.  In particular, the level of  atmospheric CO$_2$ concentration consistent with the first occurrence of an ice-free Arctic ($SIA_t = 0$) occurs when \textit{CO2C}$_{t}   = -a / b$.

The ``univariate'' representation (\ref{equ:co2_uni}) has typically been used to fit and forecast a single sea-ice indicator -- either sea-ice area or extent. We generalize this analysis to consider thickness and volume as well.  More importantly, we also introduce a joint modeling strategy that allows two or more sea-ice indicators to be modeled together. Such a representation combines indicators using a simultaneous zero-ice constraint. For example, for two indicators, the ``bivariate'' linear carbon trend model with a ``simultaneous first IFA" constraint is given by

\begin{equation}
  \label{equ:co2_bi}
  	\begin{split}
	  x_t = a^{x} +  b^{x} ~ \text{\textit{CO2C}}_{t} + \varepsilon^{x}_t \;  \\
	  y_t = a^{y} +  \frac{a^{y} b^{x}}{a^{x}} ~ \text{\textit{CO2C}}_{t} + \varepsilon^{y}_t \; ,
	  \end{split}
\end{equation}
where $x_t$ and $y_t$ are two sea-ice indicators. This regression model has jointly constrained slopes and intercepts so that $x_t = y_t = 0$ at the same level of \textit{CO2C}$_{t}$ and can be estimated via non-linear least squares.

Like the univariate equation  (\ref{equ:co2_uni}), the bivariate system (\ref{equ:co2_bi}) can be used to characterize a first (simultaneous) IFA in terms of \textit{CO2C} -- that is, a concentration level for a seasonal ice-free Arctic.
 
 \label{dimension}   For our bivariate empirical analysis with (\ref{equ:co2_bi}), we always set $x_t$ to $SIA_t$ and $y_t$ to one of the other sea-ice indicators ($SIE_t,~SIT_t,$ or $SIV_t$).\footnote{\label{subtle_1}Of course, the bivariate first-IFA-constrained modeling approach can be generalized to more than two indicators. Indeed, we have applied it to 3- and 4-variable sets of our sea-ice indicators and obtained similar empirical results to those reported below.  Subtleties also arise with 3- and 4-variable systems, such as additional potentially relevant restrictions among indicators, as when, for example, $SIV$ is at least the approximate product of $SIE$ and $SIT$.}  That is, our focus here is on the three bivariate combinations: $SIA {+} SIE$, $SIA {+} SIT$, and $SIA {+} SIV$. Each indicator  pair brings additional information to the estimation of the first IFA year by incorporating additional indicators with $SIA$, which provides a common benchmark for comparing results.  In addition, each pair has  idiosyncratic characteristics that provide useful differentiation. The $SIA {+} SIE$ pair is the only one that is based on data that are essentially directly observed, and these two indicators also share a clear-cut zero-ice constraint by definition. However, this pair is limited to only sea-ice coverage, while the other combinations account for thickness as well.  The $SIA {+} SIT$ pair integrates the two key elements necessary for measuring sea ice, but again, $SIT$ is a model-processed measure that has greater measurement error. The $SIA {+} SIV$ pair combines area with the most comprehensive measure of sea ice, but again, at a likely cost of less measurement accuracy. 
  
  \label{pure} It is  worth  highlighting three  aspects of our approach.  The first is our use of  pure-trend bivariate models with no lags. In our view, such models are well suited to our dataset. Our use of annual (September) sea-ice indicators renders greater dynamic generality unnecessary. For example, the first autocorrelation of detrended September $SIE$ is only 0.06. Also, degrees of freedom are scarce in our sample of roughly 40 annual observations. An alternative vector autoregression (VAR) specification would gain little from adding dynamics at the cost of a profligate parameterization; a 4-variable VAR(3), for example, would have more than 50 parameters.
  
The second  aspect of our approach is  use of fixed deterministic carbon trends with no breaks.  We intentionally skirt the  unit-root vs.~trend-break minefield.  That  debate remains unresolved in macroeconomics, for example, after more than half a century of research, as standard data simply are not very informative about the hypotheses of interest.\footnote{\label{ref2aa}Classic macroeconomic work on stochastic vs. deterministic trend, with or without trend breaks,  includes \cite{NelsonPlosser1982} and \cite{Perron1989}, among many others;  \cite{Stock1994}  provides a thorough survey. Climate economics work grappling with the same issues includes \cite{Kaufmanetal2006a} and \cite{PerronEstrada2019}, among many others.}  Nevertheless, one can interpret our carbon trend regressions as cointegrating regressions under a unit-root worldview, and our focus on fixed carbon trends as a simple and accurate empirical approximation, as per the fit described below in Figure \ref{fig:CO2_IceCarbonzzz}.\label{stoc}

\label{linear} The third  aspect of our approach is use of \textit{linear} carbon trends, which might be violated by episodes of prolonged natural variation or by nonlinear shifts or tipping points in the ice-carbon linkage. Our stochastic modeling methodology accounts for natural variation, although its specification may be imperfect. Natural variation is emphasized in \cite{MillerNam2020} as a potential source of a future slowdown in Arctic ice melting. However, the ``hiatus'' in global warming of 1998-2012 had a minor effect on the pace of Arctic melting, which may reflect its potential status as measurement error. Alternatively, as sea-ice melt depends importantly on polar conditions, Arctic amplification of temperature may mask such temporary global fluctuations.  

Regarding nonlinearities, following the earlier literature, we are comfortable with linear carbon trends as a suitable empirical approximation, especially in light of the excellent fit evident in Figure \ref{fig:CO2_IceCarbonzzz} and discussed below.  Although many potential factors could produce a nonlinear ice-carbon relationship, linearity turns out to be a remarkably robust approximation for the entire satellite-era observational record, regardless of whether CO$_2$ is measured as cumulative emissions or atmospheric concentration (e.g., \citealp{eisenman2009nonlinear,  tietsche2011recovery, winton2011climate, NS2016, NotzSIMIP2020}). Moreover, the  empirical climate-science evidence on linearity of the ice-carbon relationship  \textit{continues to accumulate}; see for example the recent work of   \cite{bennedsen2023multivariate}, who work in a very general nonlinear state space environment, allowing for nonlinear carbon trend, but find no need for nonlinearity. The authoritative  recent review of \cite{wang2023mechanisms} accurately summarizes the state of current knowledge:
\begin{quote}
	{The literature points toward a linear, predictable response of summer sea ice extent in response to greenhouse gas emissions, despite large inter-annual variability in weather patterns, rather than an abrupt transition to seasonally ice-free conditions (monthly average sea ice area of less than 1 million km 2) consistent with tipping point
		behavior   (p. 44)}.
\end{quote}
That is,  tipping points are unlikely to affect  Arctic sea ice carbon-trend linearity, even if they  affect other aspects of climate change. 

%Related, projections based on climate models typically find that linear carbon trends remain  good out-of-sample approximations (e.g., \citealp{NS2016, NotzSIMIP2020}) at least until first NIFA.  \label{comfortable} 
 
 Of course, moving forward, the effects of diminished albedo, greater storm-generated wave action, or permafrost melt may notably accelerate sea-ice loss, or, alternatively, greater cloud cover in a warming planet may retard that loss. There are several subtleties regarding a potential melting permafrost tipping point (e.g., \citealp{Vaksetal2020}).  Permafrost degradation can occur as gradual top down thaw or as abrupt collapse of thawing soil.  Such melting permafrost releases both methane and CO$_2$.  The exact timing and proportion of these releases is uncertain, but to the extent that CO$_2$ and non-CO$_2$ greenhouse gases (GHGs) remain in a similar proportion, our linear carbon trends will still hold approximately.  The CO$_2$ emissions time trajectory and the sea-ice time trends, however, would change. A permafrost melt that quickens the pace of warming strengthens our thesis that Arctic sea ice is melting more quickly than widely projected. 

In closing this subsection, we note  that the theoretical  climate science literature has also gravitated toward linearity, as summarized by  \cite{Hillebrand2023}, precisely because of the compelling empirical evidence discussed above.\footnote{The basic theoretical idea underlying linearity is the insight that {climate system response fluxes} may be approximated  linearly, while still  reproducing  many aspects of the historical climate record -- including but not at all limited to carbon-trend linearity, as in the classic work of \cite{raupach2013exponential}.} Important linear modeling of carbon cycle land and ocean sinks, for example, includes not only the classic work of \cite{bacastow1973changes} but also the  more-recent work of  \cite{raupach2013exponential}. Similarly, important linear modeling of atmosphere/ocean energy balance ranges from  the classic work of  \cite{budyko1969effect} and \cite{sellers1969global}, to  more recent work that  introduces two-layer energy-balance models with linear heat transfer between atmosphere/ocean surface and deep-ocean layers, such as \cite{gregory2000vertical} and \cite{held2010probing}.

%, consider
%$$\frac{dx}{dt}=f(t)+\Phi(x),$$
%where $x(t)$ is the state vector of matter and energy stores (e.g., sea ice), $f(t)$ is the forcing flux vector with first element $f_1 = CO2$, and $\Phi(x)$ denotes system response fluxes. Linearization produces  $\Phi(x) = -Kx$, so that we have 
%$$\frac{dx}{dt} = f(t) - Kx,$$
%where
%$$x(t)=\int_{0}^{t} G(t{-}s)f(s)ds,$$
%and $G(t)$ is Green's function, $G(t)=\exp(-Kt).$ Hence
%$$\frac{x_{i}(t)}{\int_{0}^t  f_{1}(0) e^{rs} ds} = \frac{x_{i'}(t)}{f_1(t)}=\sum_m \frac{a_{i1}^{(m)} f_{1}(0)}{r+\lambda^{(m)}}+O\left(e^{-(r+\lambda^{(m)})}\right),$$
%where $\lambda^{(m)}$ is the eigenvalue corresponding to the $m$-th eigenmode $\frac{f_{1}(0)}{r+\lambda(m)}e^{rt}$. It follows that $\frac{x_i(t)}{x_j(t)}$ and $\frac{x_{i'}(t)}{f_1(t)}$ converge to fixed ratios, consistent with linear trend.

\subsection{Carbon-Trend Fits and Forecasts in Ice-Carbon Space} \label{ICspace}

\begin{figure}[tp]
	\caption{September Arctic Sea-Ice Indicators: \\ Estimated Linear Carbon Trends in Ice-Carbon Space Under $SSP3 ~7.0$}
	\begin{center}
		
		\begin{minipage}{0.5\textwidth}
			\begin{center}
				Unconstrained Univariate Models
			\end{center}
		\end{minipage}%
		\begin{minipage}{0.5\textwidth}
			\begin{center}
				Constrained Bivariate Models
			\end{center}
		\end{minipage}
		
		\begin{minipage}{0.5\textwidth}	
			\centering
			\includegraphics[trim={0mm 00mm 0mm 00mm},clip,scale=.125]{{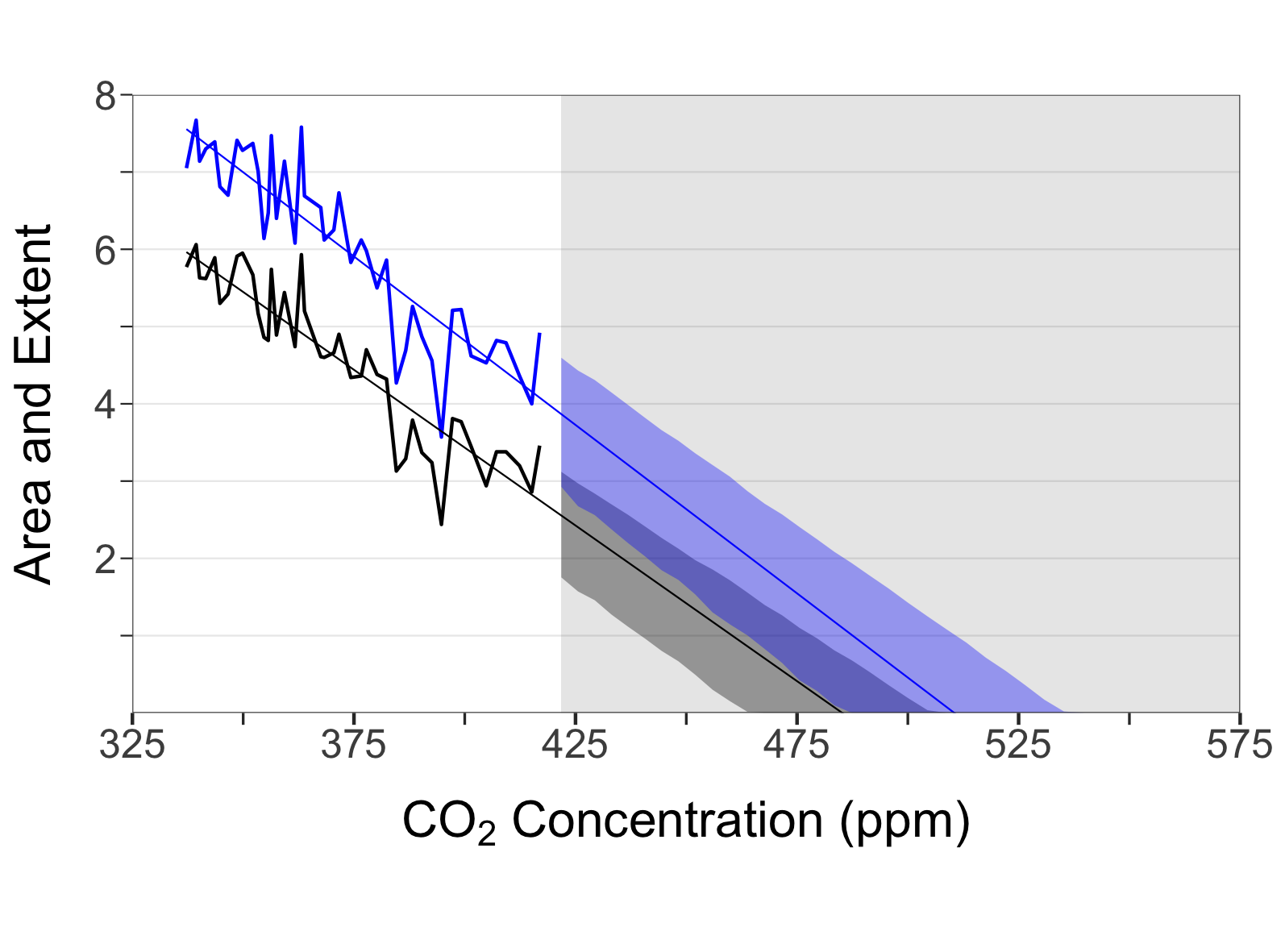}}
		\end{minipage}%
		\begin{minipage}{0.5\textwidth}	
			\centering
			\includegraphics[trim={0mm 00mm 0mm 00mm},clip,scale=.125]{{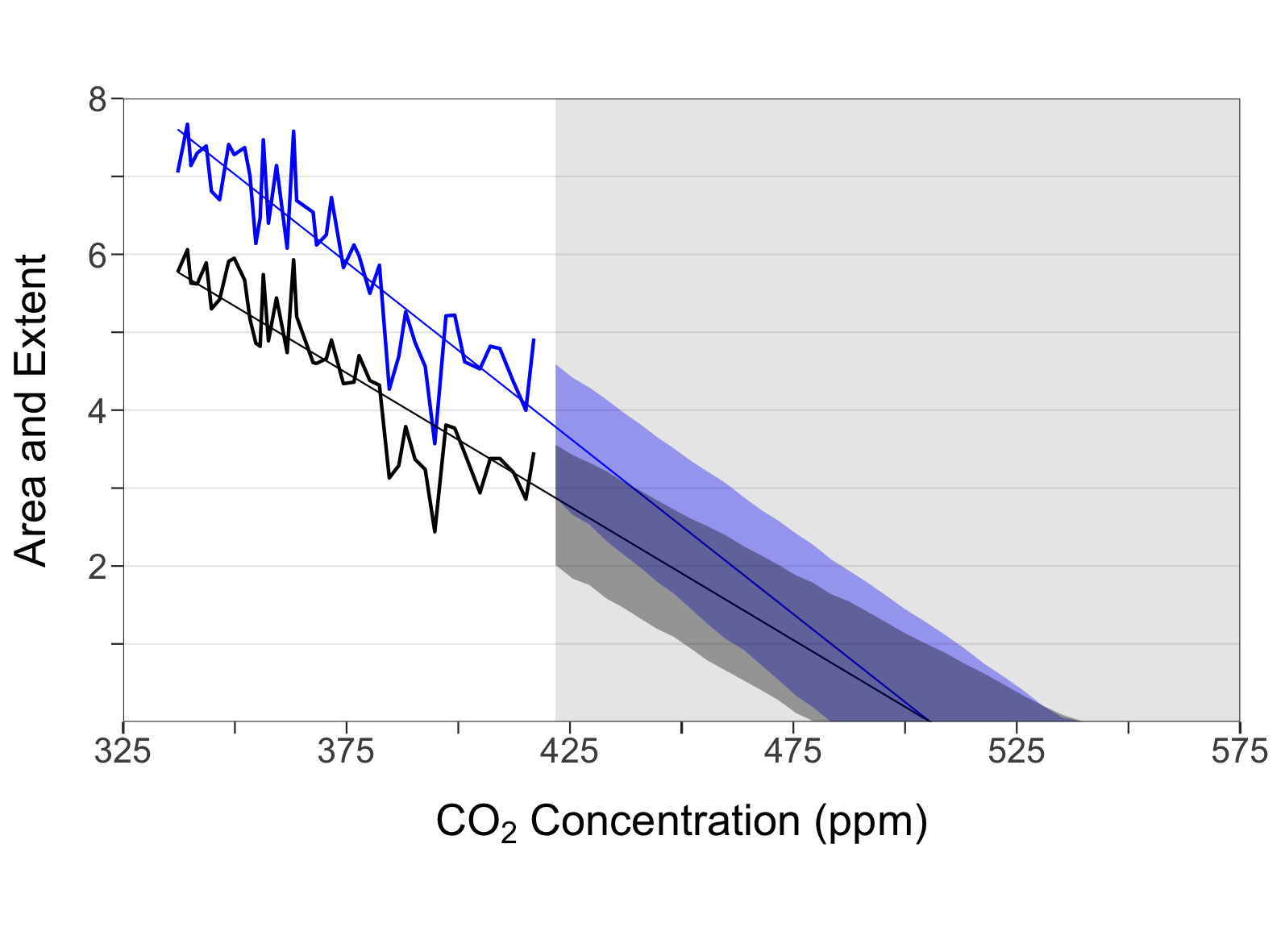}}
		\end{minipage}
		
		\begin{minipage}{0.5\textwidth}	
			\centering
			\includegraphics[trim={0mm 00mm 0mm 00mm},clip,scale=.125]{{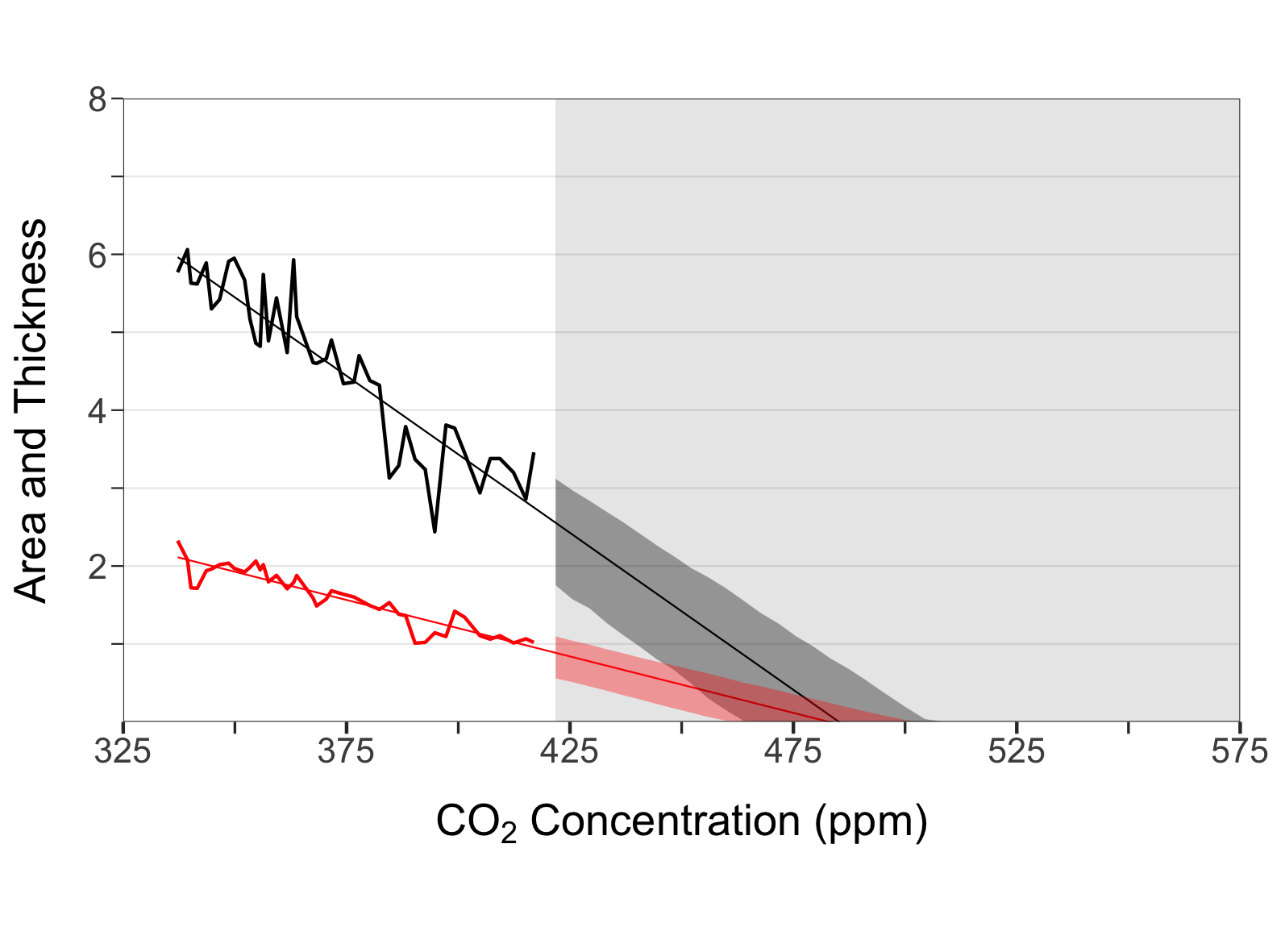}}
		\end{minipage}%
		\begin{minipage}{0.5\textwidth}	
			\centering
			\includegraphics[trim={0mm 00mm 0mm 00mm},clip,scale=.125]{{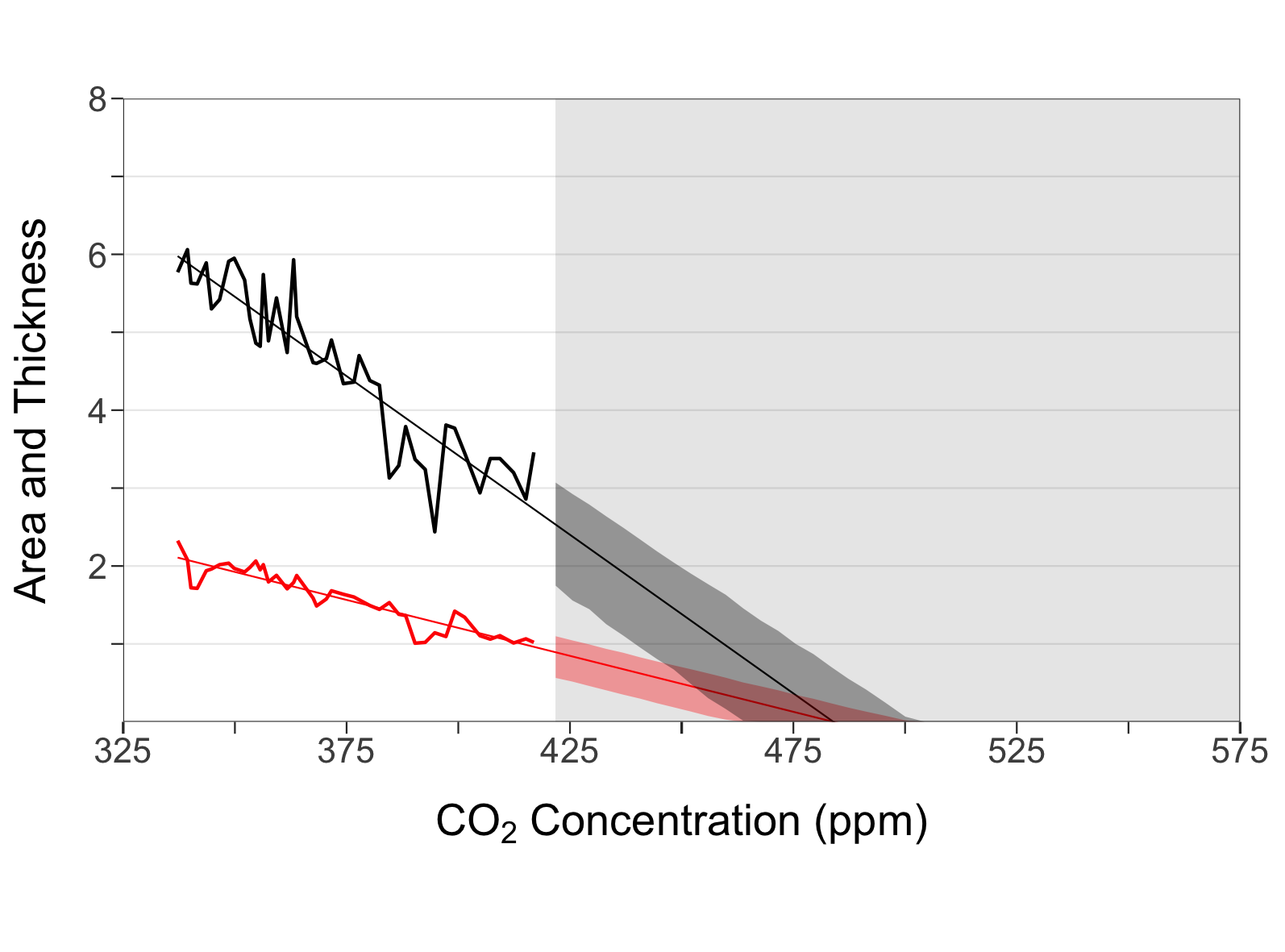}}	
		\end{minipage}
		
		\begin{minipage}{0.5\textwidth}	
			\centering
			\includegraphics[trim={0mm 00mm 0mm 00mm},clip,scale=.125]{{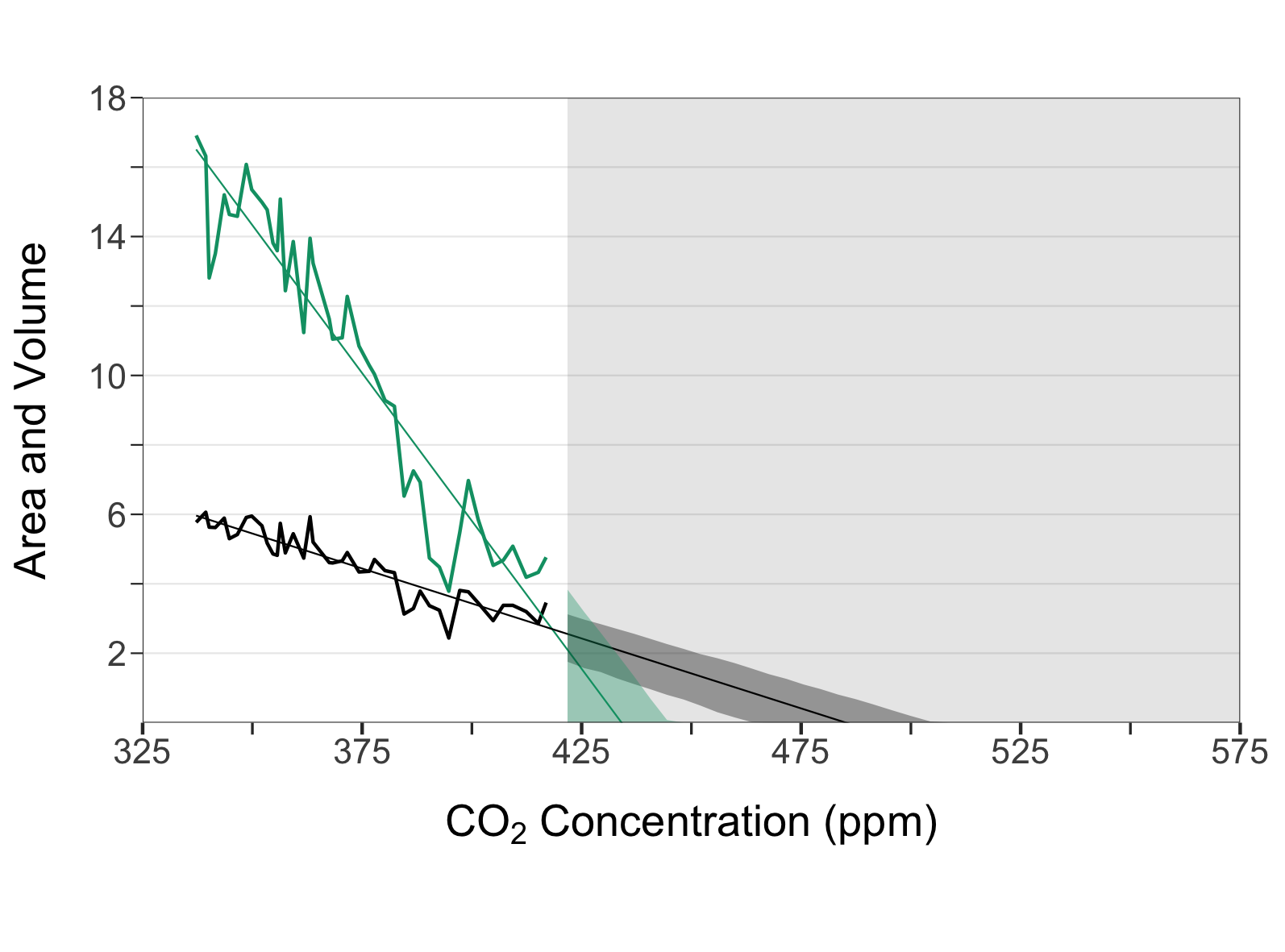}}		
		\end{minipage}%
		\begin{minipage}{0.5\textwidth}	
			\centering
			\includegraphics[trim={0mm 00mm 0mm 00mm},clip,scale=.125]{{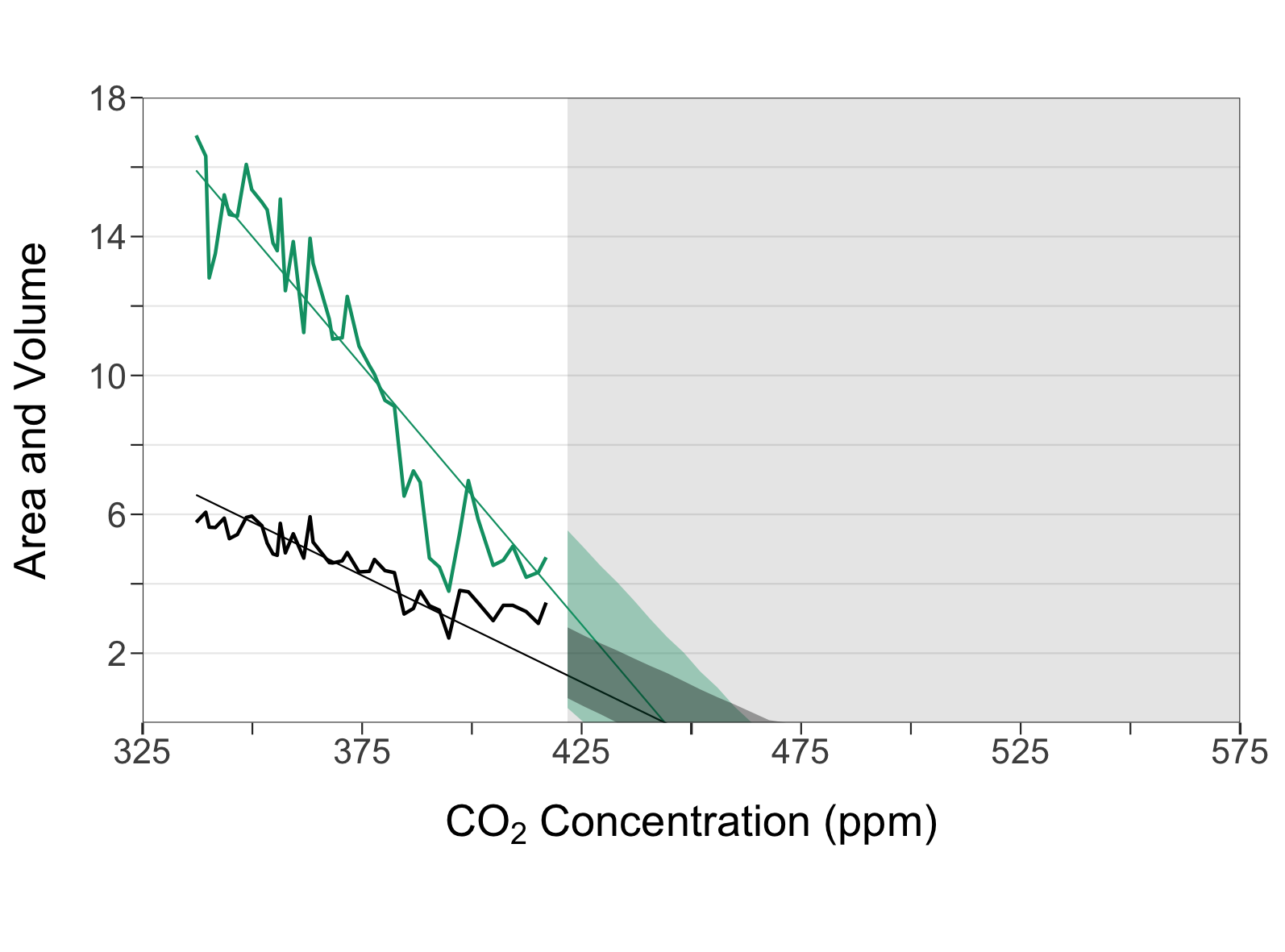}}
		\end{minipage}
	\end{center}
	\label{fig:CO2_IceCarbonzzz}
	\begin{spacing}{1.0} \noindent  \footnotesize Notes:  Each panel displays a pair of sea-ice indicators graphed against \textit{CO2C} (atmospheric CO$_{2}$ concentration, measured in ppm).  We show extent in blue (measured in $10^6\, {\rm km}^2$),  area in black (measured in $10^6\, {\rm km}^2$), thickness in red (measured in m), and volume in green (measured in $10^3\, {\rm km}^3$).  	In the univariate column, we show linear carbon-trend fits and projections based on Equation \eqref{equ:co2_uni}. 	In the bivariate column, we show linear carbon-trend regression fits and forecasts constrained to reach zero simultaneously, based on Equation \eqref{equ:co2_bi}. The historical sample period (unshaded) is 1979-2021, and the projection intervals obtained by simulation have 90\% coverage.  See Appendix \ref{bootstrap} for details.
	\end{spacing}
\end{figure}

\begin{table} [tbp]
	\caption{September Arctic Sea-Ice Indicators: \\
		Linear CO$_2$ Carbon Trend Estimates and Projections (Under SSP3 7.0)}\label{tab:CO2}\ \\
	\vspace*{-1.1cm}
	\begin{center}

		\begin{tabular}{l rrcrcr}
			\toprule  \addlinespace[8pt]
			
			& &  & \multicolumn{2}{c}{First IFA} & \multicolumn{2}{c}{First NIFA (Area)} \\
			\cmidrule(lr){4-5} \cmidrule(lr){6-7} 
			
			\multicolumn{1}{c}{Model}  & 
			\multicolumn{1}{c}{$\hat{b}$} &\multicolumn{1}{c}{$R^2$} & 
			\multicolumn{1}{c}{\textit{CO2C} (ppm)} & \multicolumn{1}{c}{year} & 
			\multicolumn{1}{c}{\textit{CO2C} (ppm)} & \multicolumn{1}{c}{year} \\
			
			\cmidrule(lr){1-7} \addlinespace[5pt]
			& \multicolumn{6}{c}{Unconstrained Univariate Models} \\
			\cmidrule(lr){2-7} \addlinespace[5pt]

			Area & 
			$\underset{(0.003)}{-0.04}$ & 0.84 &
			488 & 2039 & 464 & 2033 	\\  \addlinespace[3pt]
			Extent & 
			$\underset{(0.003)}{-0.04}$ & 0.80 &
			514 & 2045 &  &  \\  \addlinespace[3pt]
			Thickness  & 
			$\underset{(0.001)}{-0.01}$ & 0.86 &
			484 & 2038 &  &  \\  \addlinespace[3pt]
			Volume  & 
			$\underset{(0.009)}{-0.17}$ & 0.90 &
			437 & 2026 &  &  \\  
			
			\cmidrule(lr){1-7} \addlinespace[5pt]
			& \multicolumn{6}{c}{Constrained Bivariate Models} \\
			\cmidrule(lr){2-7} \addlinespace[5pt]

			Area & 
			\multirow{2}{*}{$\underset{(0.003)}{-0.03}$} & \multirow{2}{*}{0.82} & 
			\multirow{2}{*}{509} & \multirow{2}{*}{2044} & 
			\multirow{2}{*}{480} & \multirow{2}{*}{2037}   \\
			Extent  & &  & & & &  \\	 \addlinespace[10pt]
			
			Area & 
			\multirow{2}{*}{$\underset{(0.002)}{-0.04}$} & \multirow{2}{*}{0.84} &
			\multirow{2}{*}{488} & \multirow{2}{*}{2039} & 
			\multirow{2}{*}{460} & \multirow{2}{*}{2032}  \\
			Thickness & &  & & & &  \\	 \addlinespace[10pt]
			
			Area  & 
			\multirow{2}{*}{$\underset{(0.003)}{-0.06}$} & \multirow{2}{*}{0.58} &
			\multirow{2}{*}{445} & \multirow{2}{*}{2028} & 
			\multirow{2}{*}{430} & \multirow{2}{*}{2024}  \\
			Volume & &  & & & &  \\	 \addlinespace[10pt]
			
			\bottomrule
		\end{tabular}
		
	\end{center}
	\begin{spacing}{1.0}  \noindent  \footnotesize  Notes:  We estimate linear \textit{CO2C} carbon-trend models for $SIA$, $SIE$, $SIT$, and $SIV$ from 1979 to 2021. The top panel shows univariate models for each indicator, and the bottom panel shows constrained bivariate models for $SIA$ paired with $SIE$, $SIT$, or $SIV$ that impose a simultaneous first IFA. We report parameter estimates, $R^2$'s,  and projected concentrations and years at first IFA and first NIFA (the latter for $SIA$ only). The first IFA and NIFA years are based on the time path under the {SSP}3 7.0 scenario. Throughout, standard errors appear in parentheses.   See text for details.  
			
	\end{spacing}
\end{table}  

Fitted linear carbon trends for all four Arctic sea-ice indicators are shown in Figure \ref{fig:CO2_IceCarbonzzz}, along with sea-ice forecasts under an assumed SSP3 7.0 carbon path in the shaded sample. 

Each panel graphs a pair of sea-ice measures against the atmospheric CO$_{2}$ concentration. Unconstrained univariate results appear in the left panels, and constrained common-IFA bivariate results for various pairs of indicators appear in the right panels.  
The black irregular lines show $SIA$ (measured in $10^6~{\rm km}^2$) versus \textit{CO2C}, and the blue, red, and green irregular lines are for  $SIE$, $SIT$, and $SIV$ (measured in $10^6~{\rm km}^2$, m, and $10^3~{\rm km}^3$, respectively). These various standard units of measurement are such that the pairs of $SIA$, $SIE$, $SIT$, and $SIV$ can be plotted conveniently together in the various panels of Figure  \ref{fig:CO2_IceCarbonzzz}, but this convenience is irrelevant for our results. Trend regression coefficient estimates depend on the choice of measurement units, but, crucially, the implied estimates of our key objects of interest, namely, the extrapolated years of the first ice-free Arctic, do not depend on these units.  

In Figure \ref{fig:CO2_IceCarbonzzz}, linear carbon trends, again colored black, blue, red, and green for $SIA$, $SIE$, $SIT$, and $SIV$, are fitted to the data in the unshaded sample. Their fit illustrates how well the linear regressions capture the relationships between the sea-ice indicators and  atmospheric CO$_2$ concentration. For each measure, the historical data cluster quite tightly around the fitted carbon trends. This remarkably robust linearity has been noted for sea-ice area in the literature (e.g., \citealp{NS2016}), and here we generalize this result to other measures.  
In percentage terms, Arctic sea-ice coverage and thickness have trended downward at a similar rate, with $SIE$, $SIA$, and $SIT$ falling by about 50\% over the sample.  When combined, these declines account for the 75\% drop in $SIV$ over the sample.

In the shaded regions of Figure \ref{fig:CO2_IceCarbonzzz}, the carbon trends are extrapolated beyond the historical sample until the arrival of an ice-free Arctic.
The unconstrained univariate column reveals  that projected carbon at first IFA is higher for $SIE$ than for $SIA$,  slightly lower for $SIT$, and notably lower for $SIV$.  
Comparing the univariate and constrained bivariate columns of Figure \ref{fig:CO2_IceCarbonzzz}, one sees that imposition of the simultaneous first-IFA constraint  changes the slopes of the fitted and extrapolated carbon trends.  This is consistent with the fact that for each pair of indicators, the common constrained first-IFA carbon levels lie between the two unconstrained first-IFA carbon levels.

The qualitative insights from Figure \ref{fig:CO2_IceCarbonzzz} are detailed in Table \ref{tab:CO2}. The estimated ${b}$'s  are all negative and highly statistically significant, and all $R^2$ values are above .80.  The  univariate projected first-IFA concentrations for  $SIA$, $SIE$, $SIT$, and $SIV$ are 488 ppm,  514 ppm,  484 ppm, and  437 ppm, respectively. When $SIA$ is modeled jointly with $SIE$, $SIT$, and $SIV$, the bivariate constrained projected first-IFA concentrations are  509 ppm,  488 ppm, and  445 ppm.  That is, the bivariate first-IFA carbon levels are higher or lower than the univariate estimate depending on which additional indicator is blended with $SIA$.  Note that the bivariate constrained blending of $SIA$ with the other indicators reduces the range of the projected first-IFA concentrations; the univariate range is  514 ppm -  437 ppm =  77 ppm, whereas the constrained bivariate range is  509 ppm -  445 ppm =  64 ppm.

While our bivariate common first-IFA constraint imposes the joint occurrence of a zero-ice event, the literature has often focused on the occurrence of a nearly ice-free Arctic, or NIFA, defined as an $SIA$ or $SIE$ level of 1 million km$^2$ or less. This definition reflects a view that certain Northern coastal regions, notably, the Canadian Arctic Archipelago, may retain small amounts of landfast sea ice even after the open Arctic Sea is ice free \citep{WangOverland2009}.  However, the resistance of such  landfast ice to melting remains an open research issue, and there is much uncertainty about how resilient such a circumscribed ``Last Ice'' refuge will be to further warming \citep{Cooley2020, Mudryk2021, Schweiger2021}. Reflecting this uncertainty, while we also consider the usual definition of a nearly ice-free Arctic ($10^6$ km$^2$) for $SIA$ and $SIE$, we do not account for any possible structural shift in near-zero sea-ice dynamics. 

Accordingly, Table \ref{tab:CO2} reports the projected first-NIFA concentrations for $SIA$. For $SIT$ and $SIV$, there are no similar definitions of nearly ice-free concentrations in the literature, so we concentrate on the occurrence of a first NIFA only in $SIA$. Of course, any first-NIFA concentration  is lower than the associated first-IFA level, but the precise difference depends on the indicator(s) examined and model used (univariate vs. bivariate). The univariate projected $SIA$ first-NIFA concentration is  464 ppm.  It almost matches the constrained bivariate $SIA$+$SIT$ $SIA$ first-NIFA projection at 460 ppm, but is still much higher than that for $SIA$+$SIV$, and lower than that for $SIA$+$SIE$.  The range of the first-NIFA concentration projections is  480 ppm -  430 ppm =  50 ppm. 

The ``carbon budget'' is a popular concept used to quantify the remaining amount of carbon that can be emitted until some climate event is reached. Analogously, we consider a carbon budget as the difference between the \emph{current}  atmospheric CO$_2$ concentration and the concentration associated with the first occurrence of IFA or NIFA. 
Given a 2021 \textit{CO2C} value of 417 ppm and an estimated concentration of 464 ppm at first NIFA (measured under the $SIA$ univariate model), our estimate of the Arctic carbon budget available until the likely occurrence of a nearly ice-free Arctic is 47 ppm. 
This concentration carbon budget is slightly higher or lower for the other model pairs.

Finally, although we have thus far discussed only point forecasts, we also provide simulation-based 90\% interval forecasts in Figure \ref{fig:CO2_IceCarbonzzz}, shown as shaded regions around the point forecasts.\footnote{\label{boot}The interval forecasts are obtained from 1000 bootstrap  simulations, and they account for both parameter estimation uncertainty and disturbance uncertainty. For details, see Appendix \ref{bootstrap}.}  The intervals  are generally quite narrow, despite their appropriate widening as the projection horizon lengthens. \label{width1} Note also that our models are constrained in expectation only; that is, their conditional mean functions are constrained but their shock variances are not, so that some intervals are always wider than others (compare, e.g., the $SIA$ and $SIV$ intervals). 

\label{qualifications} \label{methane} There are of course many under-appreciated or unmodeled aspects of Arctic sea-ice diminution that may affect the accuracy of both our point and interval forecasts.  Some of these may speed or slow sea-ice loss.
They include exceptional internal variation, which may result in a warming hiatus  \citep{MillerNam2020} that slows loss, and tipping points associated with methane release from permafrost melting  \citep{Chylek2022}, which could promote an even steeper decline in Arctic sea ice. 
        
\subsection{Carbon-Trend Fits and Forecasts in Ice-Time Space}  \label{icetime9}

\begin{figure}[tp]
	\caption{September Arctic Sea-Ice Indicators: \\ Estimated Linear Carbon Trends in Ice-Time Space Under $SSP3 ~7.0$}
	\begin{center}
		
		\begin{minipage}{0.5\textwidth}
			\begin{center}
				Unconstrained Univariate Models
			\end{center}
		\end{minipage}%
		\begin{minipage}{0.5\textwidth}
			\begin{center}
			Constrained	Bivariate Models
			\end{center}
		\end{minipage}
		
		\begin{minipage}{0.5\textwidth}
			\centering
			\includegraphics[trim={0mm 00mm 0mm 00mm},clip,scale=.125]{{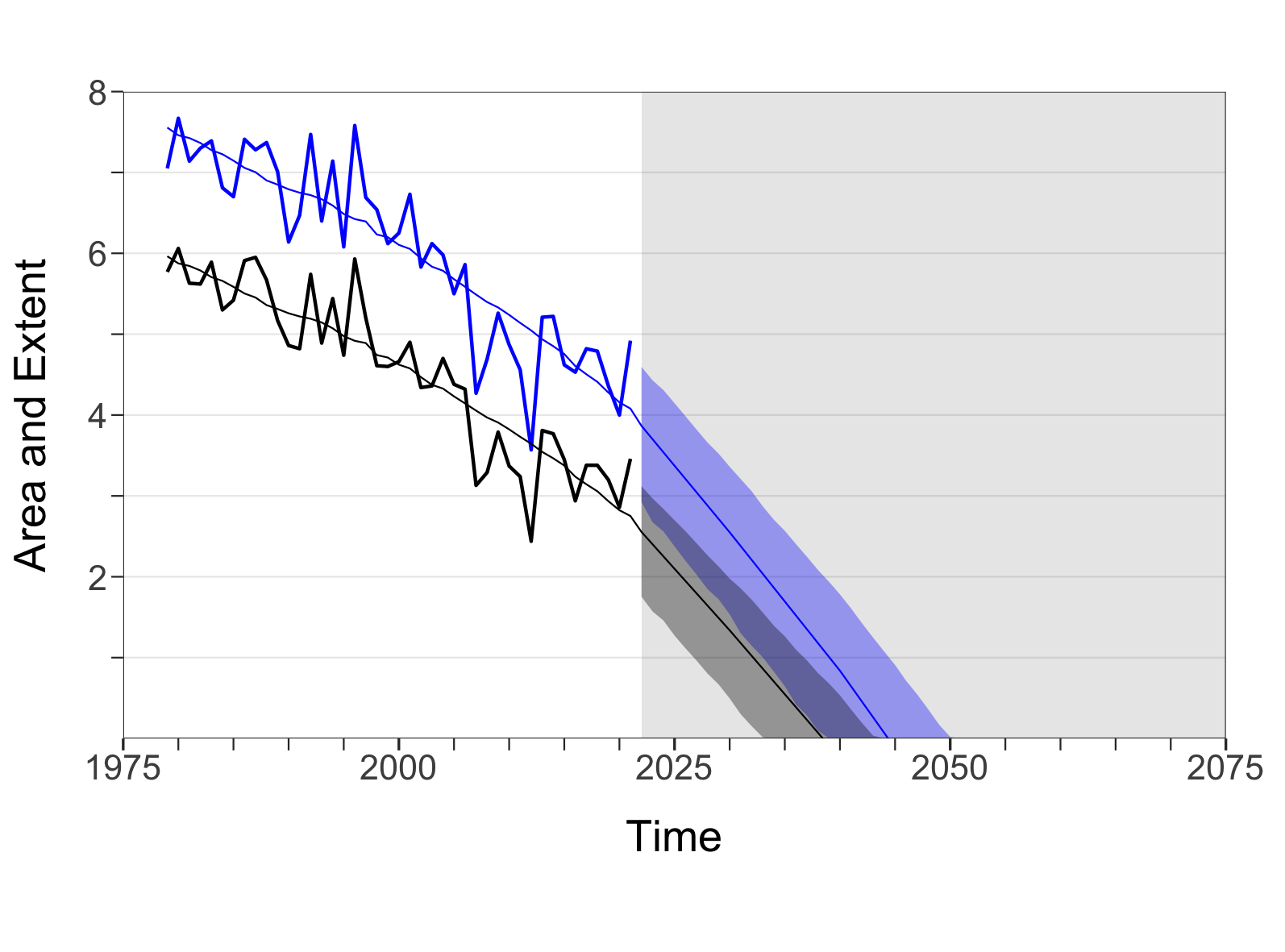}}
		\end{minipage}%
		\begin{minipage}{0.5\textwidth}
			\centering
			\includegraphics[trim={0mm 00mm 0mm 00mm},clip,scale=.125]{{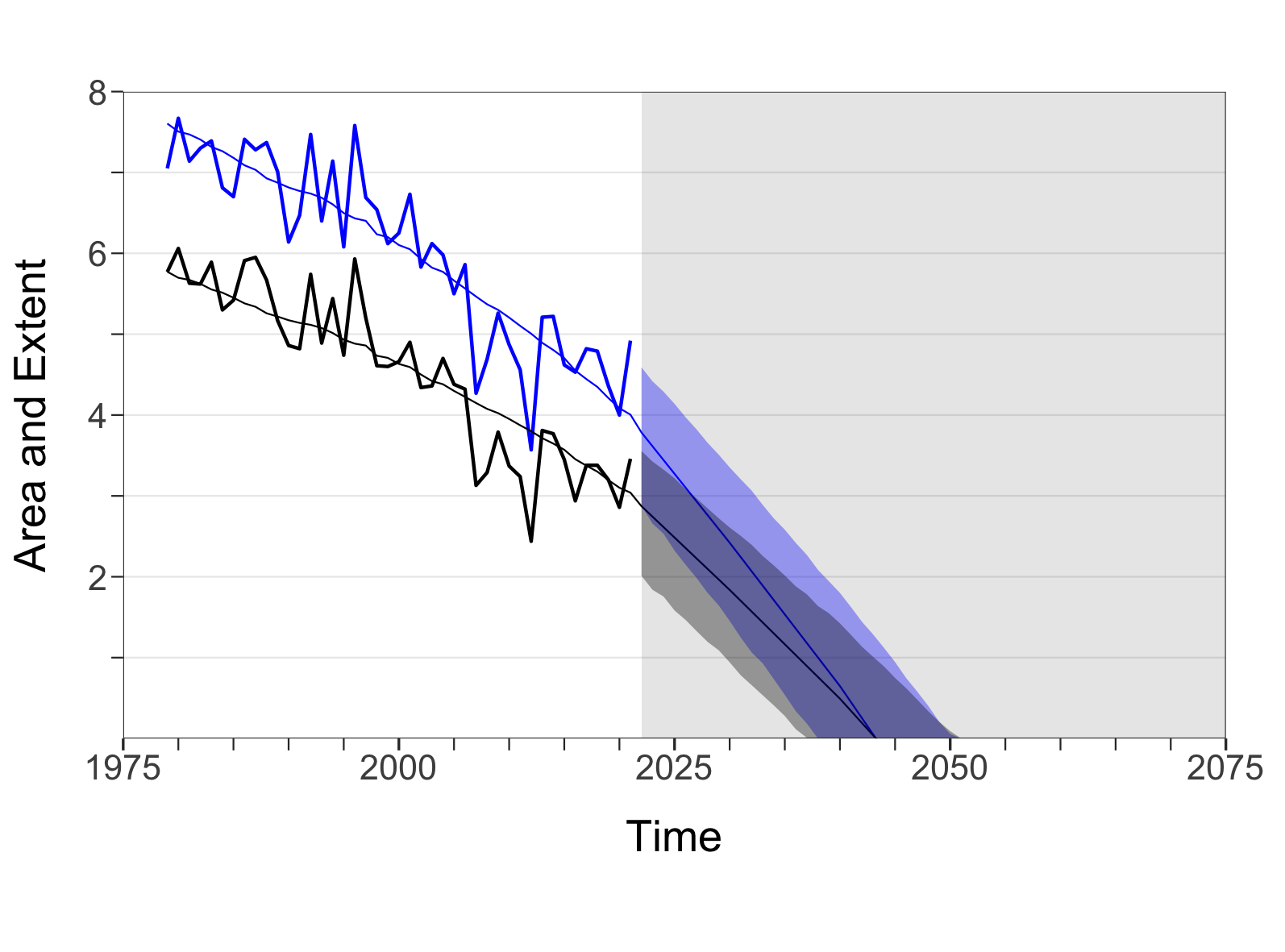}}
		\end{minipage}
		
		\begin{minipage}{0.5\textwidth}
			\centering
			\includegraphics[trim={0mm 00mm 0mm 00mm},clip,scale=.125]{{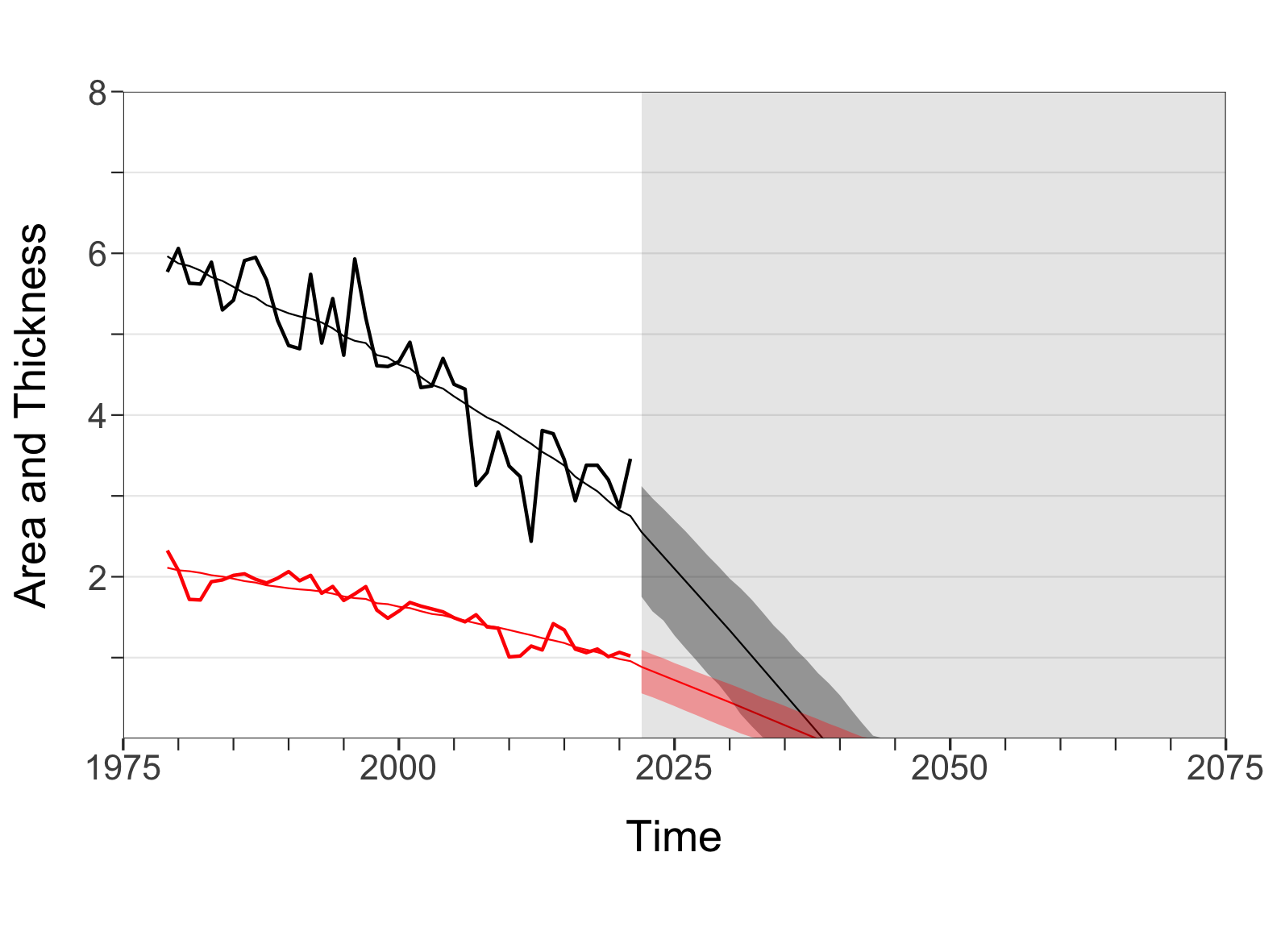}}
		\end{minipage}%
		\begin{minipage}{0.5\textwidth}
			\centering
			\includegraphics[trim={0mm 00mm 0mm 00mm},clip,scale=.125]{{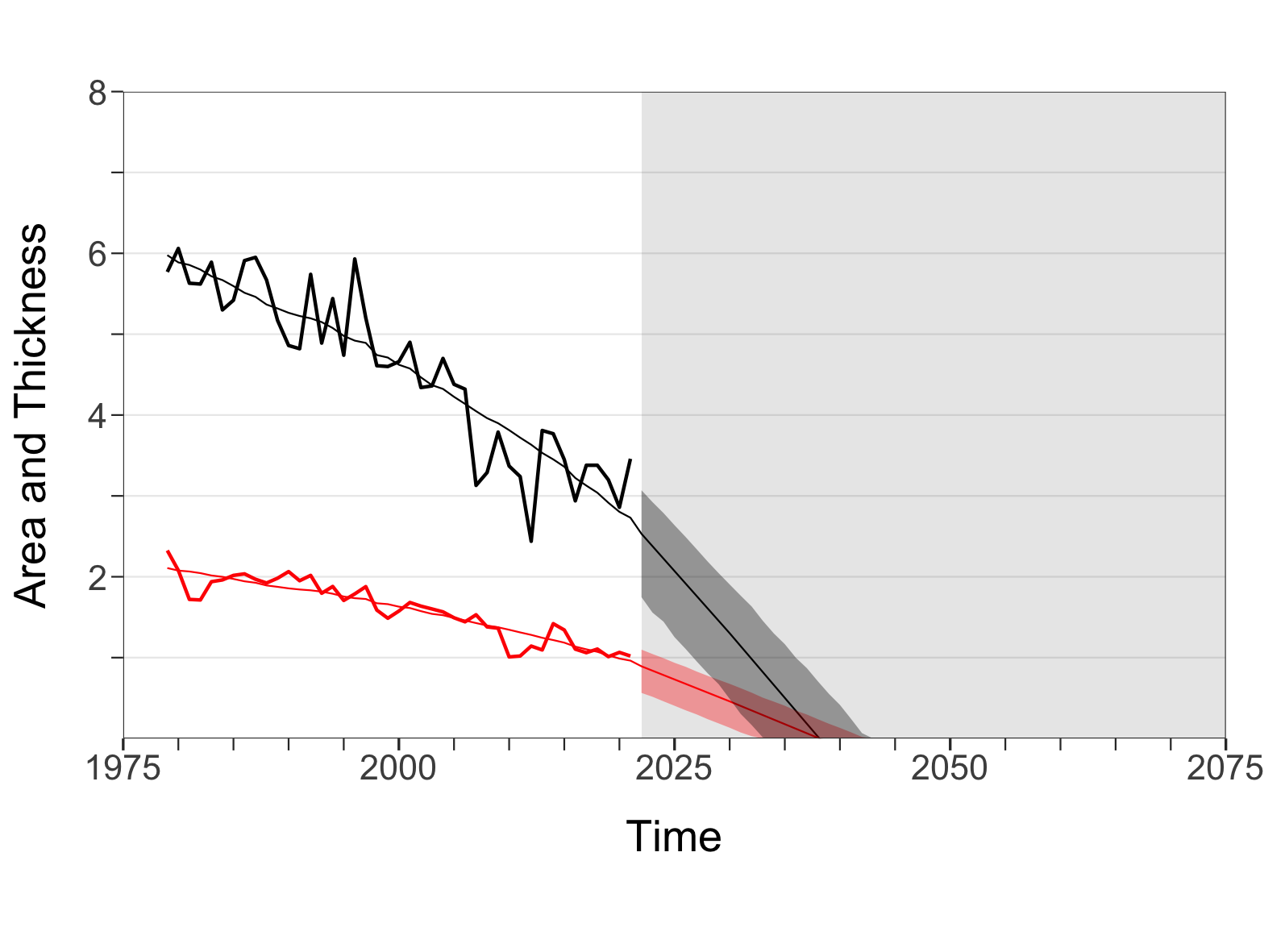}}	
		\end{minipage}
		
		\begin{minipage}{0.5\textwidth}
			\centering
			\includegraphics[trim={0mm 00mm 0mm 00mm},clip,scale=.125]{{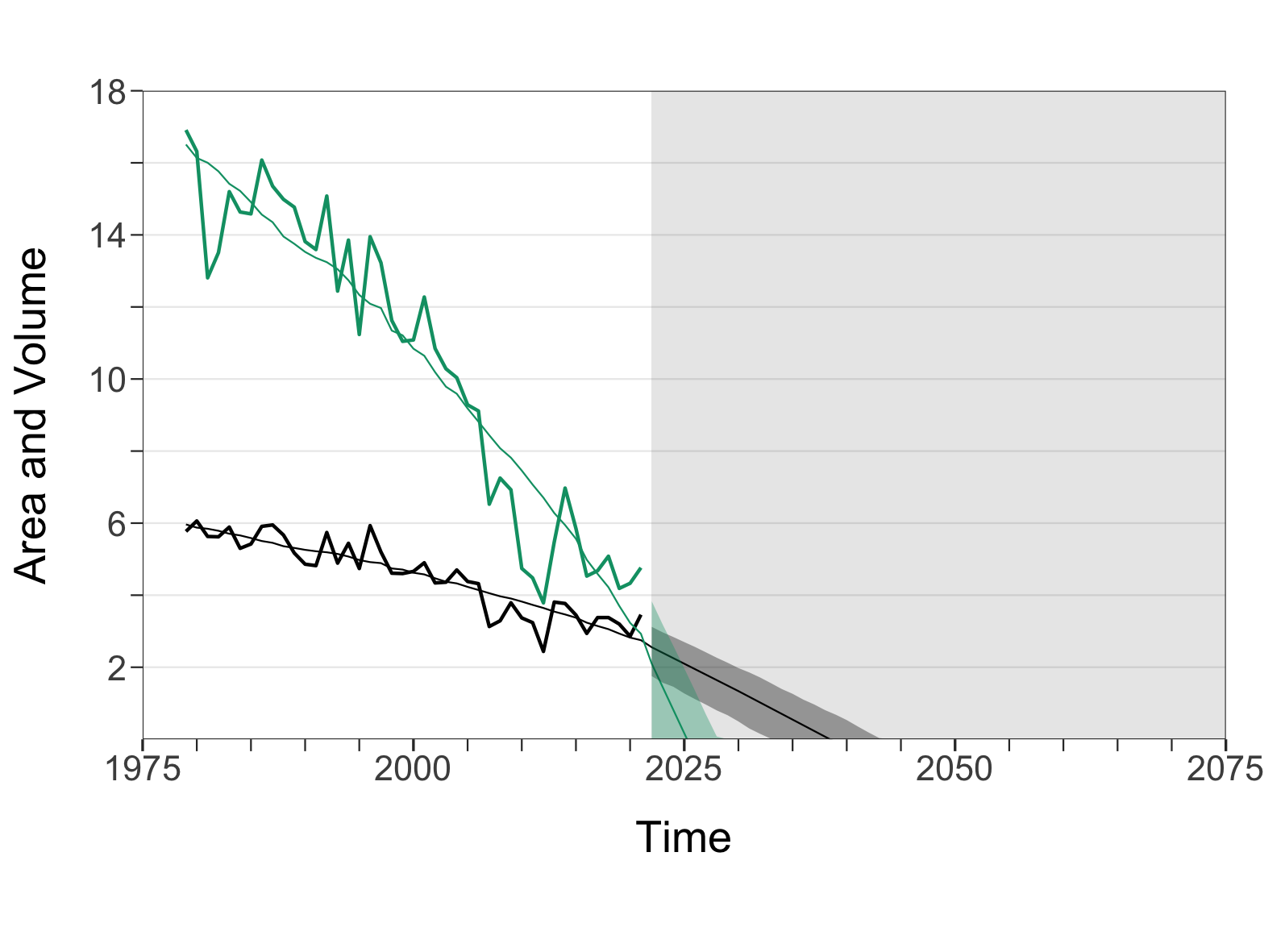}}
		\end{minipage}%
		\begin{minipage}{0.5\textwidth}	
			\centering	
			\includegraphics[trim={0mm 00mm 0mm 00mm},clip,scale=.125]{{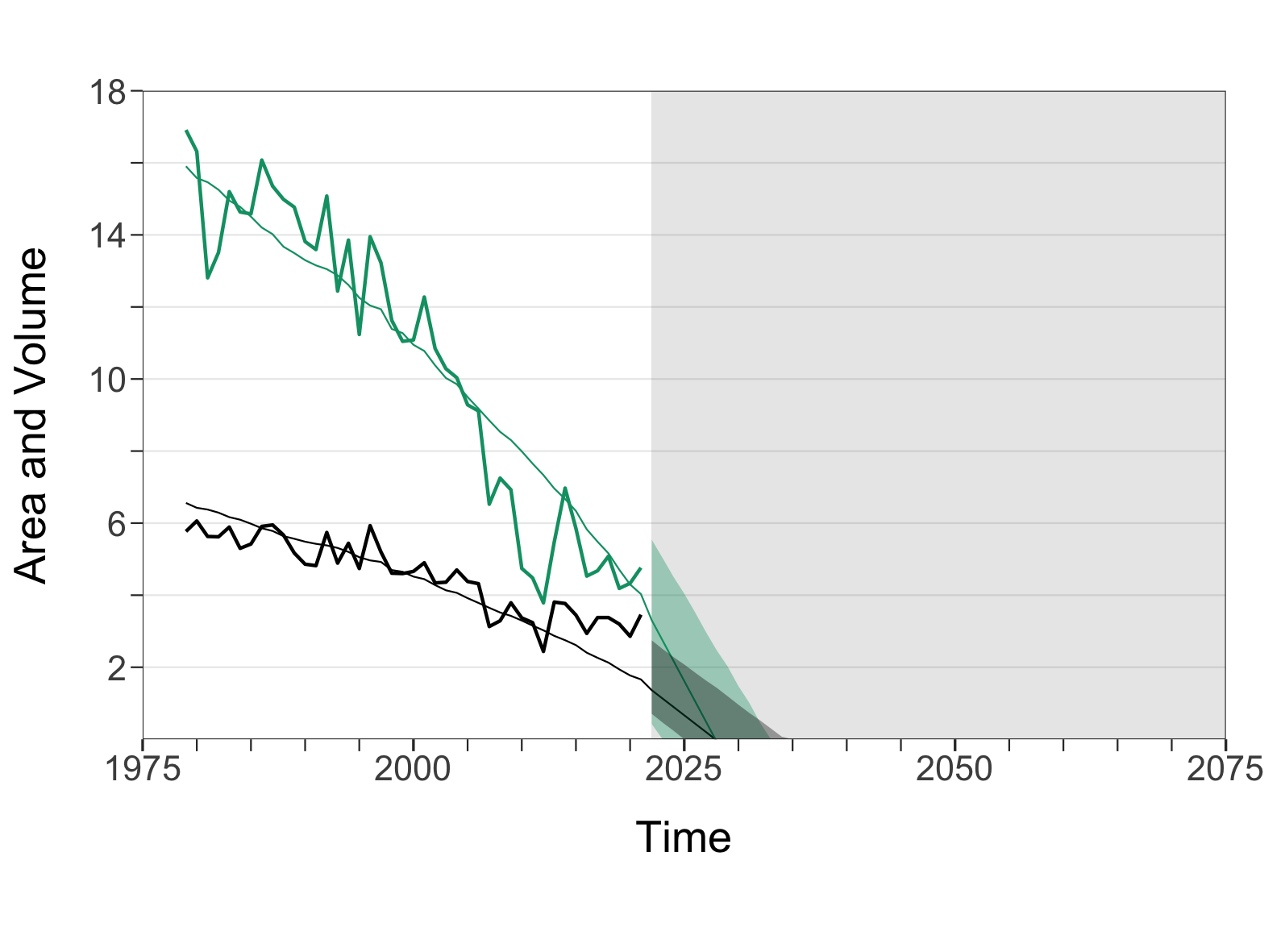}}
		\end{minipage}
	\end{center}
	
	\label{fig:CO2_IceTimezzz}
	\begin{spacing}{1.0} \noindent \footnotesize Notes: Each panel displays a pair of sea-ice indicators graphed against time.   We show extent in blue (measured in $10^6\, {\rm km}^2$),  area in black (measured in $10^6\, {\rm km}^2$), thickness in red (measured in m), and volume in green (measured in $10^6\, {\rm km}^3$).  		In the univariate column, we show linear carbon-trend fits and forecasts  based on Equation \eqref{equ:co2_uni}) converted from ice-carbon space to ice-time space under SSP3 7.0.   In the bivariate column, we show linear carbon-trend regression fits and forecasts  constrained to reach zero simultaneously, based on Equation \eqref{equ:co2_bi} converted from ice-carbon space to ice-time space under SSP3 7.0. The underlying CO$_2$ measure is atmospheric concentration, measured in ppm.  The historical sample period (unshaded) is 1979-2021, and the projection intervals obtained by simulation have 90\% coverage.  See Appendix \ref{bootstrap} for details.
	\end{spacing}
\end{figure}

The monotonically increasing  CO$_2$ concentrations in the historical data and SSP projected scenarios of Figure \ref{fig:emissions} allow us to translate carbon-trend representations  like (\ref{equ:co2_uni}) (in ice-carbon space) into time-trend representations (in ice-time space).  In particular, the one-to-one SSP scenario mappings  allow us to translate the first IFA/NIFA carbon level projections into projected first IFA/NIFA \emph{years}, conditional on the SSP scenario.  

Moreover, the   curvature of the {SSP} scenarios produces a curvature in the time trends implied by the linear carbon trends.  Because $SIA$ is linearly related to \textit{CO2C} and the {SSP} scenarios used to convert carbon trends into time trends are nonlinear, the implied $SIA$ time trend will be nonlinear.  For example, with the  CO$_2$ concentration in the SSP 7.0 scenario increasing over time at an increasing rate, the implied sea-ice time trend will be decreasing at an increasing rate.
\label{conditioning} Of course, translating the carbon-trend forecasts into ice-time space requires a second layer of conditioning relative to the carbon-trend forecasts in ice-carbon space insofar as inverse SSP scenarios are used to infer ice-time profiles, including NIFA and IFA times.  That is, the timing of the first NIFA predictions from a given estimated carbon-trend model is conditional on the assumption that future emissions follow a certain SSP path.

The implied fitted and forecasted nonlinear {time} trends are shown in Figure  \ref{fig:CO2_IceTimezzz}, the format of which parallels that of Figure \ref{fig:CO2_IceCarbonzzz}.  The implied time trends  are obtained by running the fitted carbon trends of Figure \ref{fig:CO2_IceCarbonzzz} through the (inverse) concentration schedule  {SSP}3 7.0 of Figure \ref{fig:emissions}.  Accordingly, each  concentration  in ice-carbon space yields a corresponding time in ice-time space.  For example, as discussed earlier, our univariate estimate of concentration at first NIFA is  464 ppm.  As seen by following the upper dashed horizontal line in Figure \ref{fig:emissions} (and as recorded in Table \ref{tab:CO2}), a concentration  of  464 ppm translates under SSP3 7.0 into a first NIFA year of 2033.

Table \ref{tab:CO2} contains implied projected first-IFA and first-NIFA years. The  univariate projected  $SIA$ first-IFA year is 2039 and the corresponding bivariate constrained projected  first-IFA  years are  2044,  2039, and  2028, depending on whether $SIA$ is modeled jointly with $SIE$, $SIT$, or $SIV$, respectively.  The range of univariate projected first-IFA years is 2045 - 2026 = 19 years, and the constrained bivariate range is 2044 - 2028 = 16 years.  Table \ref{tab:CO2} also reports projected $SIA$ first-NIFA  years. The univariate projected $SIA$ first-NIFA  year is 2033.  This date is close to the constrained bivariate projected $SIA$+$SIT$ first-NIFA year and ranges between those for $SIA$+$SIV$ and $SIA$+$SIE$, respectively.  The range of the projected first-NIFA years is 2037 - 2024 = 13 years. Across all univariate and bivariate specifications, the average $SIA$ first-NIFA data falls between 2031 (average bivariate) and 2033 (univariate), so the occurrence of a nearly ice-free Arctic is clearly projected to occur fairly soon. 
  
Finally, as we will describe in detail below, our first-IFA and first-NIFA dates are robust to use of  low, medium, or high  concentration growth scenarios.   We have emphasized the medium (SSP3 7.0) scenario as a plausible baseline, but  low (SSP2 4.5) and high (SSP5 8.5) scenarios provide similar dates, because first NIFA occurs relatively rapidly before the carbon levels in the various scenarios can diverge.  This can be seen  in Figure \ref{fig:emissions}, where the horizontal $\rm  CO_2$ concentration line at 464 ppm  cuts the medium and high  SSP schedules at nearly-identical times, and the low SSP schedule just a few years later.  That is, our results indicate that global adoption of a somewhat lower emissions path would likely delay the arrival of a seasonally ice-free Arctic by only a few years.   

\section{Quadratic Time-Trend Models, Fits, and Forecasts}  \label{trendpro}

\begin{figure}[tp]
	\caption{September Arctic Sea-Ice Indicators: \\ Estimated Quadratic Time Trends}
	\begin{center}
		
		\begin{minipage}{0.5\textwidth}
			\begin{center}
				Unconstrained Univariate Models
			\end{center}
		\end{minipage}%
		\begin{minipage}{0.5\textwidth}
			\begin{center}
				Constrained	Bivariate Models
			\end{center}
		\end{minipage}
		
		\begin{minipage}{0.5\textwidth}	
			\centering
			\includegraphics[trim={0mm 00mm 0mm 00mm},clip,scale=.125]{{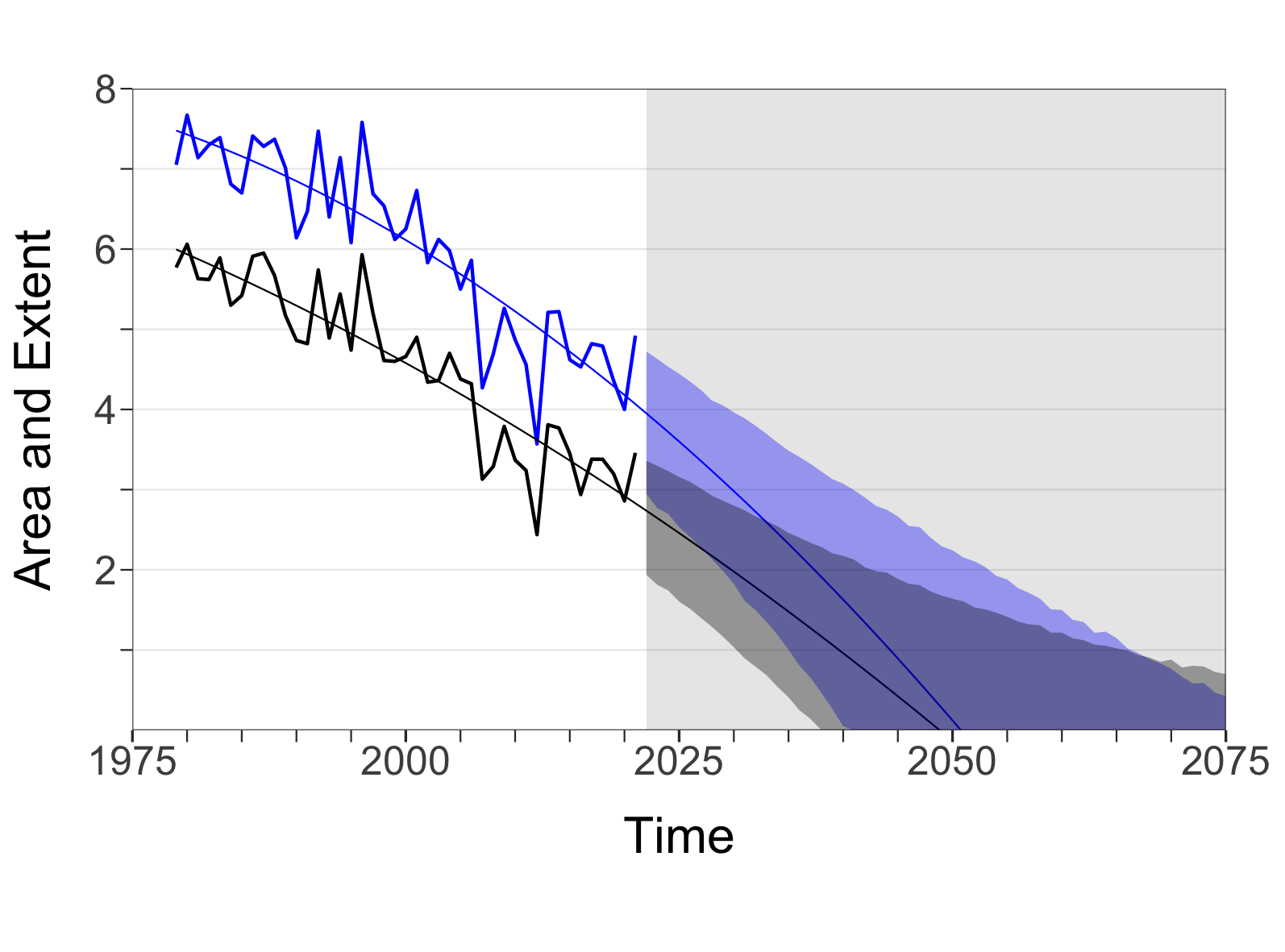}}
		\end{minipage}%
		\begin{minipage}{0.5\textwidth}	
			\centering
			\includegraphics[trim={0mm 00mm 0mm 00mm},clip,scale=.125]{{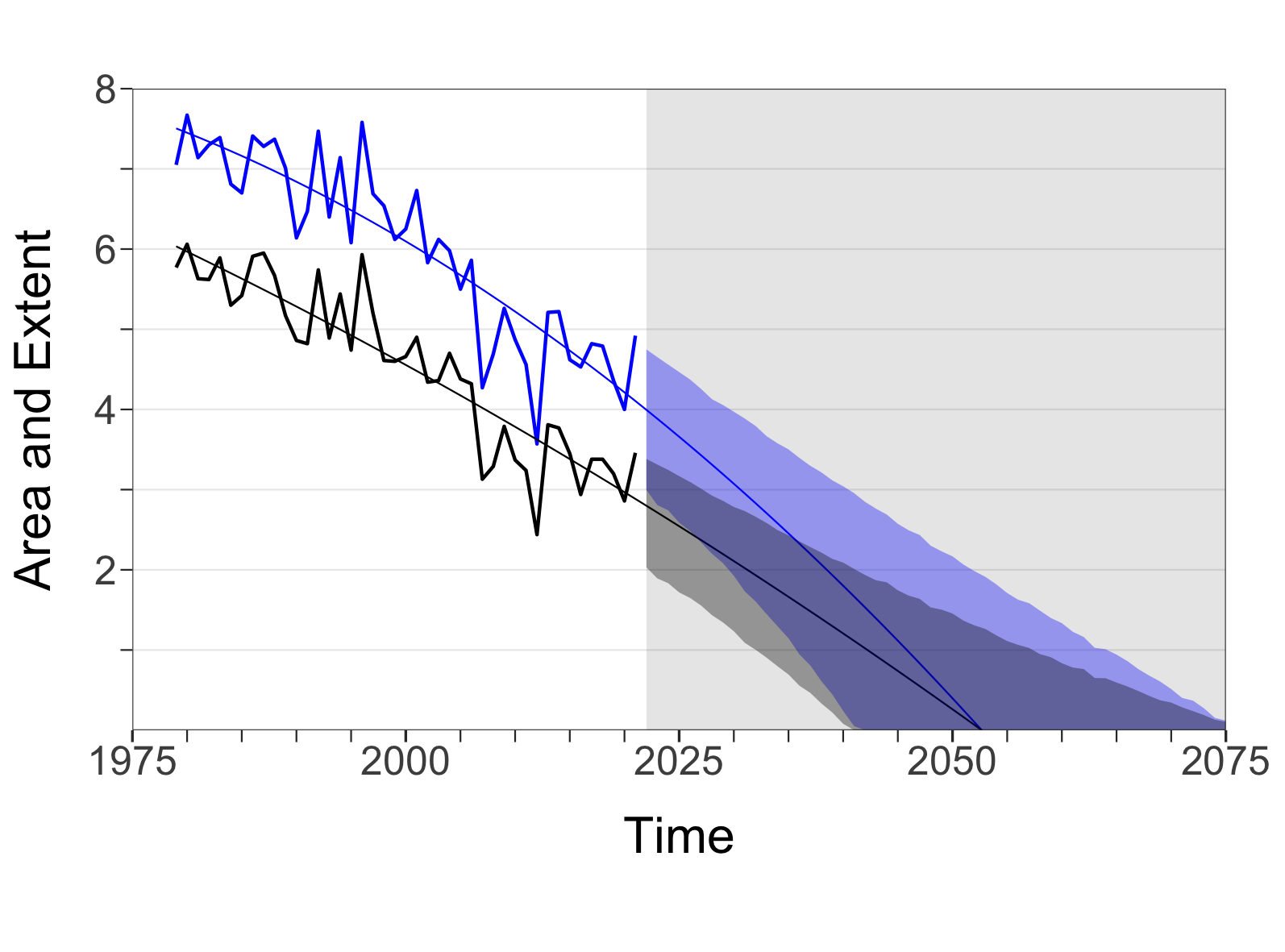}}
		\end{minipage}
		
		\begin{minipage}{0.5\textwidth}	
			\centering
			\includegraphics[trim={0mm 00mm 0mm 00mm},clip,scale=.125]{{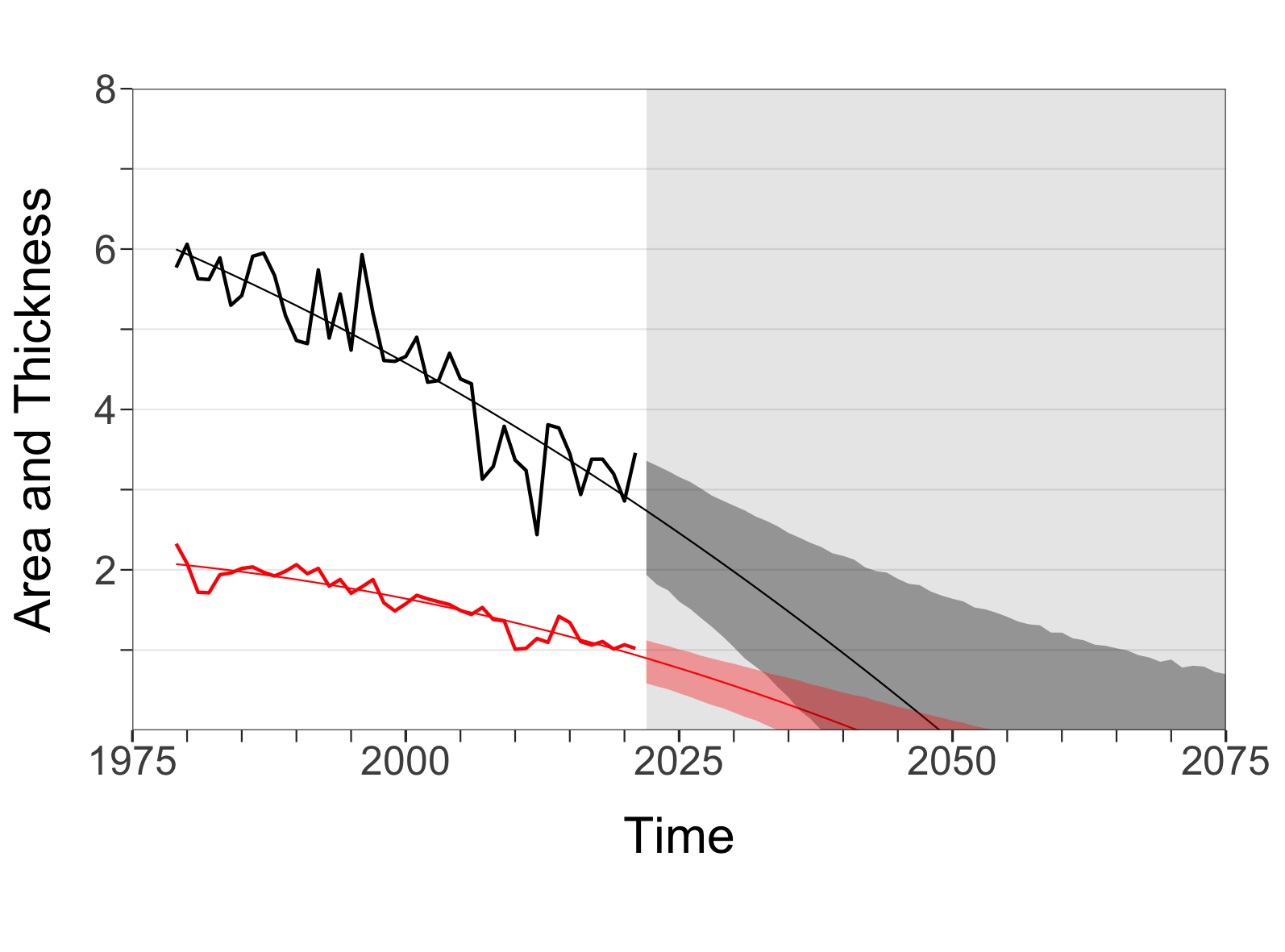}}
		\end{minipage}%
		\begin{minipage}{0.5\textwidth}	
			\centering
			\includegraphics[trim={0mm 00mm 0mm 00mm},clip,scale=.125]{{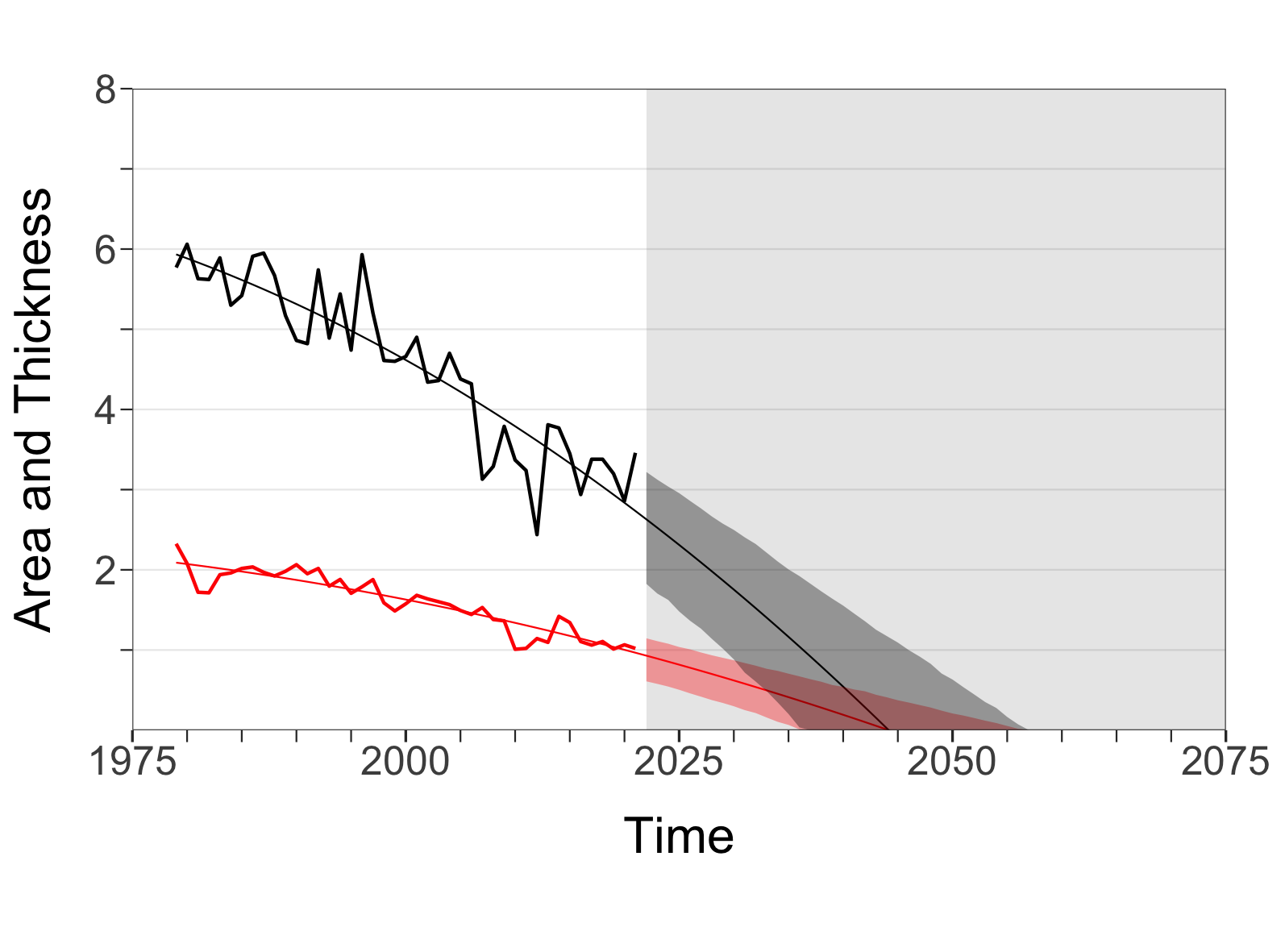}}	
		\end{minipage}
		
		\begin{minipage}{0.5\textwidth}	
			\centering
			\includegraphics[trim={0mm 00mm 0mm 00mm},clip,scale=.125]{{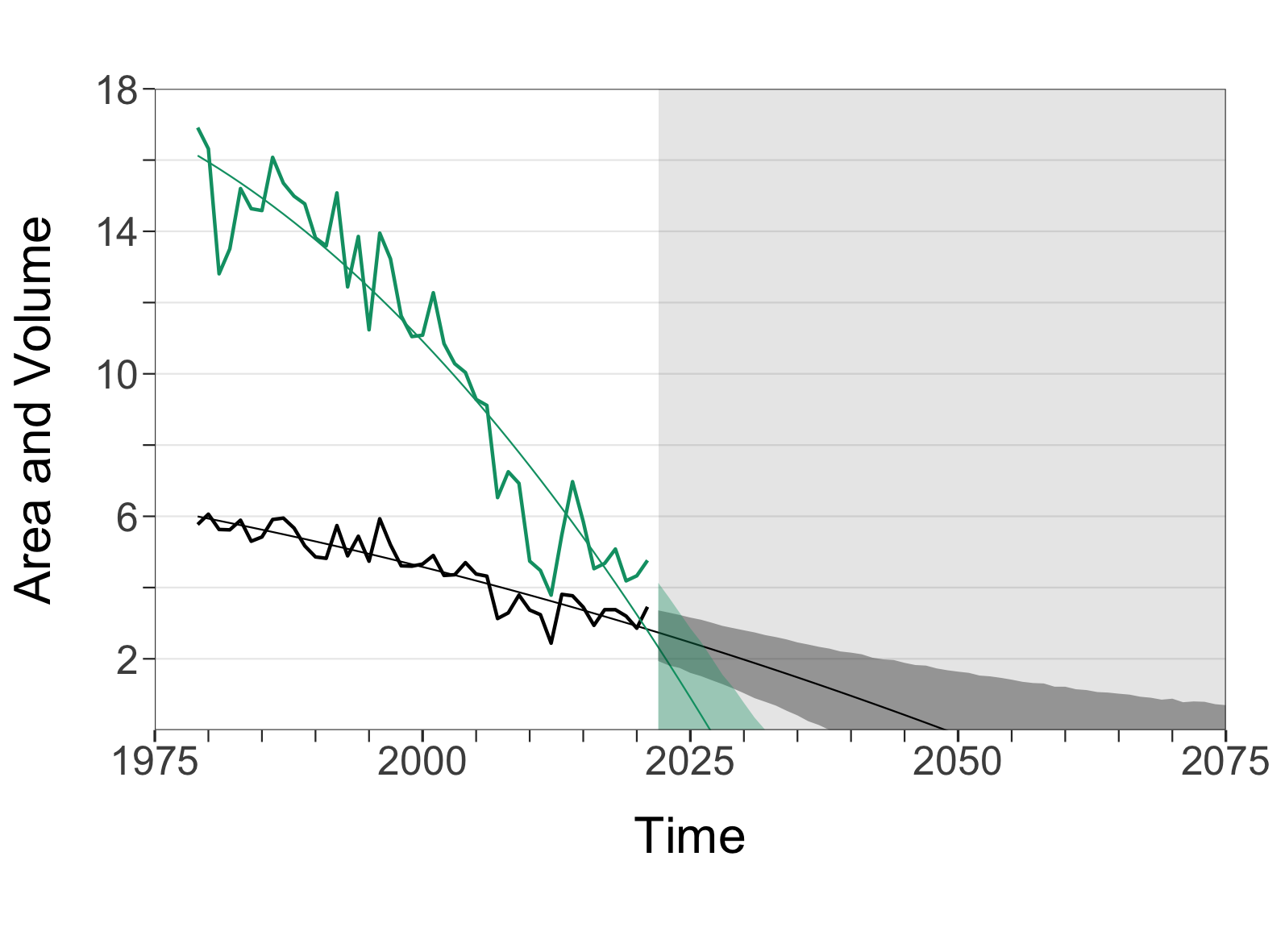}}
		\end{minipage}%
		\begin{minipage}{0.5\textwidth}	
			\centering		
			\includegraphics[trim={0mm 00mm 0mm 00mm},clip,scale=.125]{{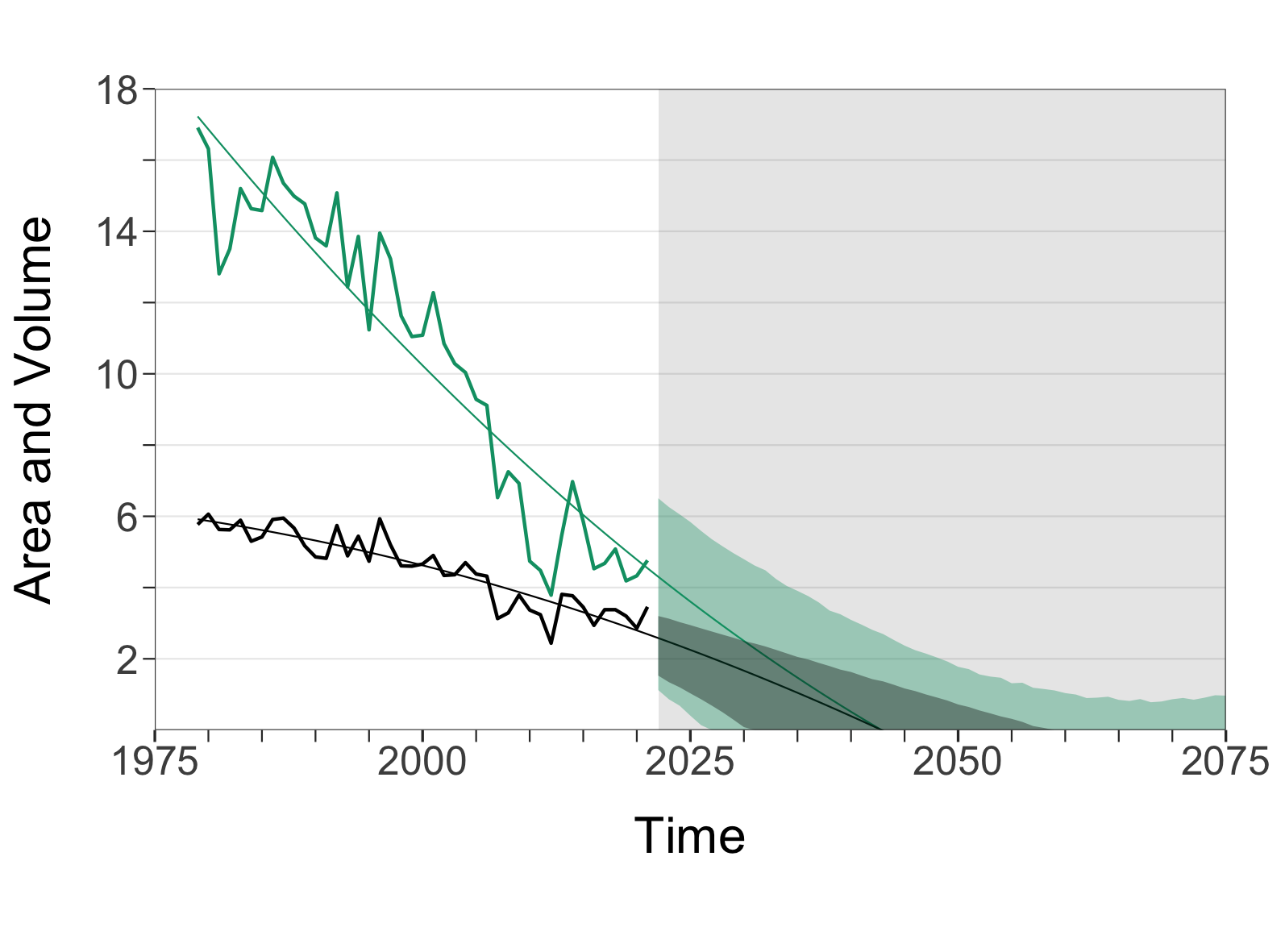}}
		\end{minipage}
	\end{center}
	\label{fig:TIME}
	\begin{spacing}{1.0}  \noindent  \footnotesize  Notes: We show extent in blue (measured in $10^6\, {\rm km}^2$),  area in black (measured in $10^6\, {\rm km}^2$), thickness in red (measured in m), and volume in green (measured in $10^6\, {\rm km}^3$).  
		In the univariate column, we show direct quadratic time-trend regression fits and forecasts, based on Equation \eqref{levelxxx}. 
		In the bivariate column, we show direct quadratic time-trend regression fits and forecasts constrained to reach zero simultaneously, based on Equation \eqref{root}. The historical sample period (unshaded) is 1979-2021, and the projection intervals obtained by simulation have 90\% coverage.  See Appendix \ref{bootstrap} for details.
	\end{spacing}
\end{figure}

In the previous section, we focused on the connection between higher atmospheric CO$_2$ concentration and melting Arctic ice, implicitly recognizing that higher greenhouse gas concentrations raise global surface temperatures and lead to melting ice. Remarkably, a simple linear regression appeared adequate, based on the past forty years, for forecasting \textit{CO2C} at first IFA and NIFA.  We then used a leading SSP concentration scenario to convert the forecasted sea-ice paths from ice-carbon space to ice-time space.  That is, we ultimately forecasted the path and pattern of $SIA$ diminution over time, and in particular, the years of first IFA and first NIFA. The nonlinear SSP path converted the linear carbon trend into a nonlinear implied time trend, decreasing at an increasing rate.
  
As a first and major robustness check, we now compare the nonlinear time trends obtained from the ice-carbon relationship to deterministic polynomial trends fit directly to $SIA$, as in \cite{DRice}.  In particular, we allow for the  relevant ice-time nonlinearity by allowing for quadratic time effects. In the univariate approach, we use

\begin{equation} \label{levelxxx}
	SIA_t = a^{x} + b^{x}  TIME_t + c^{x}  TIME_t^2 + \varepsilon^{x}_{t}
\end{equation}
where $SIA_t$ is sea-ice area,  $TIME$ is a time trend dummy, and $\epsilon_t$ represents deviations from  trend.  Similarly, in the  bivariate approach we use
\begin{equation*} \label{level}
	x_t = a^{x} + b^{x}  TIME_t + c^{x}  TIME_t^2 + \varepsilon^{x}_{t}
\end{equation*}
\begin{equation*}
	y_t = a^{y} + b^{y}  TIME_t + c^{y}  TIME_t^2 +  \varepsilon^{y}_{t},
\end{equation*}
where $x_t$ is $SIA_t$ and $y_t$ is  one of the other sea-ice indicators ($SIE_t,~SIT_t, ~SIV_t$).  Linearity of course emerges as a special case.

In our flagship constrained bivariate approach, we move to the equivalent root form of the quadratic system,
\begin{equation}\label{root}
	x_t = (\alpha^{x} - \beta^{x} \, TIME_t)  \,  (1 - \gamma^{x}  \,  TIME_t)  + \varepsilon^{x}_{t}
\end{equation}
\begin{equation*}
	y_t = (\alpha^{y} - \beta^{y}  TIME_t)  \,  (1 - \gamma^{y}  \,  TIME_t) + \varepsilon^{y}_{t},
\end{equation*}
where the roots of the $x$ trend are $1 / \gamma^{x}$ and $\alpha^{x} / \beta^{x}$, and the roots of the $y$ trend are $1 / \gamma^{y}$ and $\alpha^{y} / \beta^{y}$. Imposing the common first-IFA  constraint (i.e., both trends reach zero simultaneously)  amounts to constraining the two quadratics to have a common root,  $\gamma^x \, {=} \, \gamma^y \, {=} \, \gamma$, which can immediately be done in a joint estimation of the two quadratic trends.\footnote{That is, in the joint regression, $a^{i} = \alpha^{i}$, $b^{i} = -\beta^{i} -\alpha^{i} \gamma$, and $c^{i} = \beta^{i} \gamma$, for $i=x,y$.  }

The fitted direct quadratic time trends are shown in Figure \ref{fig:TIME}.  The point forecast paths are generally similar to those of the carbon-implied time trends in Figure \ref{fig:CO2_IceTimezzz}.  The interval forecast paths are wider, however, than for carbon-implied time trends.  This is to be expected because the carbon-implied time trends of Figure \ref{fig:CO2_IceTimezzz} \textit{assume} the amount of curvature via an assumed SSP path on which they condition, whereas the direct time trends of Figure \ref{fig:TIME} \textit{estimate} the curvature, which is challenging.  The extra parameter estimation uncertainty then translates into wider projection intervals.  \label{width2} Additionally,  note that   only part of a time trend is constrained by imposing identical IFA times (one root is left free),  and, as a result,  different indicators approach IFA at different speeds.  The different curvatures produce different asymmetries in the IFA intervals, with flatter trends (like $SIA$)  promoting asymmetry and steeper trends (like $SIV$) promoting symmetry. 

\label{quadratic} Put differently, the carbon-implied trends provide  forecasts {conditional} on a linear carbon-trend model and a particular assumed SSP scenario, 
whereas the directly-fitted quadratic time trends do not, instead (implicitly) conditioning simply on continued ``business as usual".  Neither approach is necessarily ``better" than the other, and indeed they both address the same  \textit{question}  (``What is the likely future path of Arctic sea ice?"), but the perspective and methods used to \textit{answer} the question differ across  the approaches.  Given the common question but different approaches to answering it, it is certainly of interest to compare the answers, and reassuring to see that they cohere closely under a leading carbon  scenario.

Parameter estimates for both the unconstrained univariate and constrained bivariate direct quadratic trends are highly statistically significant, and all $R^2$'s are above 0.80.  The key bivariate projected first-NIFA years are 2043, 2037, and 2036, for $SIA$+$SIE$, $SIA$+$SIT$, and $SIA$+$SIV$, respectively.   The projected first-NIFA year range is 2043 - 2036 = 7 years.  The mean projected first-NIFA year is 2039%\footnote{The exact value of $2038.\bar{6}$ is rounded up to 2039.}
, about 8 years after NIFA projected by the carbon-trends estimation. 

All told, the point forecasts of first-NIFA years from the carbon-trend models and the  direct time-trend models agree quite closely, in terms of NIFA arriving very soon, but the carbon-trend models all feature somewhat earlier NIFA.  We now proceed to provide a more complete characterization of the uncertainty surrounding the first-NIFA point forecasts, via a full accounting of first-NIFA probability distributions.  

\section{Probabilistic Assessment of Sea-Ice Disappearance} \label{Disappear}

\begin{table} [tbp]
	\caption{September Arctic Sea-Ice Indicators:\\Probability Distributions of $SIA$ First September NIFA Years} \ \\
	\vspace*{-1.1cm}
	\label{tab:Distributions2}
	\begin{center}
		%\footnotesize
		
		\begin{tabular}{l cccccccccc}
			\toprule  \addlinespace[8pt]
			
			& Mean & Median & Mode &  Std &  Skew & Kurt  & 5\% & 20\% & 80\% & 95\% \\
			\cmidrule(lr){2-11} \addlinespace[5pt]
						
			\addlinespace[6pt]
			\underline{SSP5  8.5}\\
			$SIA$+$SIE$ & 2034 & 2034 & 2035 & 3.20 & -0.24 & 3.03 & 2028 & 2031 & 2036 & 2039 \\
			$SIA$+$SIT$ & 2030 & 2031 & 2031 & 2.33 & -0.53 & 3.10 & 2026 & 2028 & 2032 & 2034 \\
			$SIA$+$SIV$ & 2024 & 2024 & 2023 & 1.60 & 0.68 & 3.16 & 2022 & 2023 & 2025 & 2027 \\ \addlinespace[10pt]
			
			\underline{SSP3 7.0}\\
			$SIA$+$SIE$ & 2034 & 2035 & 2035 & 3.62 & -0.11 & 2.88 & 2028 & 2031 & 2037 & 2040 \\
			$SIA$+$SIT$ & 2031 & 2031 & 2031 & 2.46 & -0.40 & 3.08 & 2026 & 2029 & 2033 & 2034 \\
			$SIA$+$SIV$ & 2024 & 2024 & 2023 & 1.63 & 0.67 & 3.14 & 2022 & 2023 & 2025 & 2027 \\ \addlinespace[10pt]
			
			\underline{SSP2  4.5}\\
			$SIA$+$SIE$  & 2039 & 2039 & 2040 & 4.98 & -0.01 & 2.76 & 2031 & 2035 & 2043 & 2047 \\
			$SIA$+$SIT$  & 2034 & 2034 & 2035 & 3.33 & -0.33 & 2.91 & 2028 & 2031 & 2037 & 2039 \\
			$SIA$+$SIV$  & 2025 & 2025 & 2024 & 2.13 & 0.64 & 3.23 & 2022 & 2023 & 2027 & 2029 \\ \addlinespace[10pt]
			
			\underline{SSP1  2.6}\\
			$SIA$+$SIE$  & 2049 & 2046 & 2094 & 14.22 & 1.48 & 5.26 & 2032 & 2038 & 2057 & 2083 \\
			$SIA$+$SIT$  & 2038 & 2038 & 2040 & 5.68 & 0.37 & 3.70 & 2029 & 2033 & 2042 & 2047 \\
			$SIA$+$SIV$  & 2025 & 2025 & 2025 & 2.44 & 0.68 & 3.35 & 2022 & 2023 & 2028 & 2030 \\ \addlinespace[10pt]
			
			\underline{Time trend}\\
			$SIA$+$SIE$  & 2039 & 2038 & 2038 & 5.49 &  0.53 & 3.57 & 2030 & 2034 & 2043 & 2048 \\
			$SIA$+$SIT$ & 2034 & 2034 & 2034 & 4.16 & 0.46 & 3.80 & 2028 & 2031 & 2038 & 2041 \\
			$SIA$+$SIV$ & 2033 & 2034 & 2035 & 5.42 & 0.07 & 2.36 & 2025 & 2028 & 2038 & 2042 \\
			\addlinespace[6pt]

			\bottomrule
			
		\end{tabular}
	\end{center}
	\begin{spacing}{1.0} \normalsize \noindent  \footnotesize Notes:  We show summary statistics for distributions of first NIFA years for linear carbon- and quadratic time-trend models. Std is standard deviation, Skew is skewness, Kurt is kurtosis, and xx\% is the xx-th percentile.  All years are rounded to the nearest integer.  For $SIA$+$SIE$ under SSP1 2.6, $SIA$ never reaches NIFA in 4.44\% of the simulations, in which case we impute the last year in which we observe NIFA for $SIA$, which is 2094.
	\end{spacing}
\end{table}

Here we maintain our focus on first-NIFA years, but we build up approximations to their full probability distributions.  Unlike our earlier first-NIFA point projections, which are based simply on extrapolated deterministic trends, here we account for random variation, which lets us learn not only about central tendencies of the first-NIFA distributions (mean, median, mode), but also other moments and related statistics (standard deviation, skewness, kurtosis, left- and right-tail percentiles, etc.) 

Results appear in Table \ref{tab:Distributions2}, based on  bivariate carbon-trend models (for  concentration scenarios SSP5 8.5, SSP3 7.0, SSP2 4.5, and SSP1 2.6) and direct time-trend models. The first three columns report the mean, median, and mode measures of central tendency of the first-NIFA year distributions.  Regardless of carbon-trend vs. direct time trend estimation, or the specific SSP scenario (high, medium, low) used with the carbon trends, or the additional indicator blended with $SIA$, it is clear that NIFA will arrive relatively soon, consistent with our first-NIFA projections based on deterministic extrapolations. The median NIFA date across the 15 median estimates in column 2 is 2034.  Only the very optimistic ``net zero by 2050" SSP1 2.6 scenario delivers an appreciable delay of a first NIFA.   

The central tendencies of the distributions of NIFA years in Table \ref{tab:Distributions2} are generally a bit \textit{earlier} than the corresponding years in Table \ref{tab:CO2}. For example, for the  leading baseline case, carbon trend $SIA${+}$SIT$ using the mid-range concentration scenario SSP3 7.0, the distribution median is 2031, a bit earlier than the 2033 reported in Table \ref{tab:CO2}.  The slightly earlier central tendencies of NIFA arrival in Table \ref{tab:Distributions} occur because, when simulating sample pathways, the addition of stochastic shocks to a declining deterministic trend, particularly one concave to the origin (such as ours), might shorten the time necessary to achieve a given threshold. The random variation raises the possibility of a transitory fall below the threshold even before the deterministic trend achieves it on a sustained basis.\footnote{Consider, for example, a gentle deterministic trend that slowly approaches zero from above. This series will continue to run just above zero for a while before  hitting zero. Adding random shocks to that trend, on the other hand, will almost certainly push the series below zero considerably sooner.  We call this the ``principle of stochastic precedence".}
According to the simulation, stochastic shocks are more likely to result in early NIFA years (79.85\%) than later NIFA years (9.51\%).  The discrepancy between the timing of the deterministic first NIFA year and the mean (or median or mode) of the stochastic first NIFA year depends on the curvature of the trend and the variance of the shocks. A steeper slope, a more concave trend curve, or a smaller error variance will reduce the timing difference.
 
The last two columns of Table \ref{tab:Distributions2} characterize the right tails of the NIFA arrival distributions.  Consider again the baseline case: carbon trend $SIA${+}$SIT$ with medium concentration scenario SSP3 7.0.  The ``80\%" column of  Table \ref{tab:Distributions2} reports an eighty percent chance of first NIFA by 2033. 
Even the most extreme late arrival (moving up and down the 80\% column) is fairly soon (2057).  Returning to the baseline case but increasing the confidence level from 80\% all the way to 95\% increases the NIFA arrival year by only one year -- from 2033 to 2034.  Alternatively, maintaining the 80\% confidence level but moving to the  optimistic  lower concentration scenario SSP2 4.5 again postpones  NIFA arrival by (only) five years.  

\section{Measuring CO$_2$ as Cumulative Emissions}  \label{emissions}

\begin{table} [tpb]
	\caption{September Arctic Sea-Ice Indicators:\\Probability Distributions of $SIA$ First September NIFA Years\\ CO$_2$ Measured as  Cumulative Emissions} \ \\
	\vspace*{-1.1cm}
	\label{tab:Distributions}
	\begin{center}
		%	\footnotesize
		
		\begin{tabular}{l cccccccccc}
			\toprule  \addlinespace[8pt]
			
			& Mean & Median & Mode &  Std &  Skew & Kurt  & 5\% & 20\% & 80\% & 95\% \\
			\cmidrule(lr){2-11} \addlinespace[5pt]

			\addlinespace[6pt]
			\underline{SSP5  8.5}\\
			$SIA$+$SIE$  & 2037 & 2037 & 2038 & 3.89 & -0.21 & 2.80 & 2030 & 2034 & 2040 & 2043 \\
			$SIA$+$SIT$  & 2033 & 2033 & 2034 & 2.80 & -0.43 & 3.00 & 2028 & 2031 & 2035 & 2037 \\
			$SIA$+$SIV$  & 2025 & 2025 & 2024 & 2.00 & 0.55 & 2.92 & 2022 & 2023 & 2027 & 2028 \\ \addlinespace[10pt]
			\underline{SSP3  7.0}\\
			$SIA$+$SIE$ & 2038 & 2038 & 2039 & 4.30 & -0.09 & 2.77 & 2030 & 2034 & 2041 & 2044 \\
			$SIA$+$SIT$ & 2033 & 2033 & 2034 & 2.98 & -0.37 & 2.92 & 2028 & 2031 & 2036 & 2038 \\
			$SIA$+$SIV$ & 2025 & 2025 & 2024 & 2.01 & 0.59 & 3.00 & 2022 & 2023 & 2027 & 2028 \\ \addlinespace[10pt]
			\underline{SSP2  4.5}\\
			$SIA$+$SIE$ & 2041 & 2041 & 2042 & 5.34 & 0.03 & 2.80 & 2032 & 2036 & 2045 & 2049 \\
			$SIA$+$SIT$ & 2035 & 2035 & 2036 & 3.57 & -0.26 & 2.78 & 2029 & 2032 & 2038 & 2040 \\
			$SIA$+$SIV$ & 2025 & 2025 & 2025 & 2.25 & 0.62 & 3.19 & 2022 & 2023 & 2027 & 2029  \\ \addlinespace[10pt]
			\underline{SSP1  2.6}\\
			$SIA$+$SIE$ & 2046 & 2045 & 2045 & 9.35 & 1.07 & 5.88 & 2033 & 2038 & 2053 & 2062 \\
			$SIA$+$SIT$ & 2038 & 2038 & 2040 & 4.92 & 0.02 & 2.76 & 2029 & 2033 & 2042 & 2046 \\
			$SIA$+$SIV$ & 2026 & 2025 & 2025 & 2.44 & 0.68 & 3.29 & 2022 & 2023 & 2028 & 2030  \\ \addlinespace[10pt]
			\underline{Time trend}\\
			$SIA$+$SIE$  & 2039 & 2038 & 2038 & 5.49 &  0.53 & 3.57 & 2030 & 2034 & 2043 & 2048 \\
			$SIA$+$SIT$ & 2034 & 2034 & 2034 & 4.16 & 0.46 & 3.80 & 2028 & 2031 & 2038 & 2041 \\
			$SIA$+$SIV$ & 2033 & 2034 & 2035 & 5.42 & 0.07 & 2.36 & 2025 & 2028 & 2038 & 2042 \\
			\addlinespace[6pt]
			\bottomrule
			
		\end{tabular}
	\end{center}
	\begin{spacing}{1.0} \normalsize \noindent \footnotesize  Notes:  We show summary statistics for distributions of first NIFA years for linear carbon-trend and quadratic time-trend constrained bivariate models. Std is standard deviation, Skew is skewness, Kurt is kurtosis, and xx\% is the xx-th percentile.  All years are rounded to the nearest integer. For $SIA$+$SIE$ under SSP1 2.6, $SIA$ never reaches NIFA in 0.16\% of the simulations, in which case we impute the last year in which we observe NIFA for $SIA$, which is 2100.
	\end{spacing}
\end{table}

In Section \ref{trendpro}, we checked the robustness of our carbon-trend results by comparing them to directly fitted time trends.  In this section, we implement a second major robustness check by providing and comparing  carbon-trend results when carbon is measured as cumulative CO$_2$ emissions since 1850 (\textit{CO2E}) rather than atmospheric CO$_2$ concentration (\textit{CO2C}).\footnote{The emissions data are based on  \cite{IPCC2021_data} and served as input to the IPCC AR6 report \citep{IPCC2021_policy}. The emissions-based regressions use the historical $CO_{2}$ data from 1979 to 2019 and two additional years of data from the {SSP}3 7.0 scenario to complete the sample. See Appendix \ref{data_app} for details.} 

%The emissions-based  carbon-trend results appear in Figure 	\ref{fig:CO2_IceCarbon} (carbon trends in ice-carbon space) and Figure 	\ref{fig:CO2_IceTime}  (implied carbon trends in ice-time space), which are quite similar to the earlier results based on concentration (compare to Figures \ref{fig:CO2_IceCarbonzzz} and \ref{fig:CO2_IceTimezzz}). A linear carbon trend for emissions appears to fit well for each measure of Arctic sea ice.  Across the panels of Figure \ref{fig:CO2_IceCarbon}, the projected cumulative emissions  at first $SIA$ IFA is generally on the order of 2842 to 4179 gt. The corresponding \textit{CO2E}  for first NIFA (univariate $SIA$) is 3275 gt, which when compared to the current cumulative emissions of about 2499 gt\footnote{{SSP}3 7.0 projection for 2021} puts the carbon budget at around 776 gt for reaching a nearly  ice-free Arctic Ocean. 
%
%The implied projected first-IFA and first-NIFA years from the emissions-based carbon-trend regressions using the SSP3 7.0 scenario are shown in Figure \ref{fig:CO2_IceTime}. The  univariate projected $SIA$ first-IFA year is 2043, and the corresponding bivariate constrained projected  first-IFA  years are 2049, 2043, and 2031, depending on whether $SIA$ is modeled jointly with $SIE$, $SIT$, or $SIV$, respectively. These are several years later than the corresponding first-IFA years obtained using atmospheric concentration.

The emissions-based  carbon-trend results  are quite similar to the earlier results based on concentration. A linear carbon trend for emissions appears to fit well for each measure of Arctic sea ice.  The projected cumulative emissions  at first $SIA$ IFA is generally on the order of 2842 to 4179 gt. The corresponding \textit{CO2E}  for first NIFA (univariate $SIA$) is 3275 gt, which when compared to the current cumulative emissions -- projected for 2021 under {SSP}3 7.0 -- of about 2499 gt puts the carbon budget at around 776 gt for reaching first NIFA. 

The implied projected first-IFA and first-NIFA years from the emissions-based carbon-trend regressions using the SSP3 7.0 scenario are also quite similar to the earlier results based on concentration.  The  univariate projected $SIA$ first-IFA year is 2043, and the corresponding bivariate constrained projected  first-IFA  years are 2049, 2043, and 2031, depending on whether $SIA$ is modeled jointly with $SIE$, $SIT$, or $SIV$, respectively. These are several years later than the corresponding first-IFA years obtained using atmospheric concentration.

The implied emissions-based first-NIFA probability distributions appear in 
Table \ref{tab:Distributions}.  They are nearly identical in shape to the earlier-reported concentration-based first-NIFA distributions, with one difference: the emissions-based distribution is shifted slightly rightward relative to the concentration-based distribution (i.e., shifted toward later first-NIFA years).  Using SSP3 7.0 and $SIA{+}SIT$, for example, produces an emissions-based median first NIFA of 2033, for example, in contrast  to our earlier  concentration-based 2031. The right-shifting of the emissions-based distribution makes its left tail thinner, so its 80th and 95th percentiles increase by even more relative to the concentration-based distribution.  Moving from concentration to emissions increases the 80th percentile  from 2033 to 2036 (again using SSP3 7.0 and $SIA{+}SIT$)  and increases the 95th percentile from 2034 to 2038.

\section{Concluding Remarks}  \label{conclsec}

We have constructed projections of September Arctic sea ice using a variety of specifications.  We prefer bivariate rather than univariate projections, because they enlarge the underlying information set in two ways.  First, the bivariate projections blend the information in the history of $SIA$ with that in the histories of other sea-ice indicators, such as $SIE$, $SIT$, and $SIV$, which may improve projection accuracy. Second, they provide regularization by enabling us to constrain first IFA to coincide across sea-ice indicators, thereby imposing appropriate geophysical constraints on otherwise flexible statistical models.

Importantly, our analysis produces full probability distribution projections of Arctic sea ice and, in particular, distributions of first-NIFA years based on constrained bivariate models.  Our empirical results reconcile linear carbon trends with quadratic time trends and provide constrained multivariate support for an early probabilistic September disappearance of sea ice.  \cite{NS2016} and others have also considered carbon regressions of the kind we use, but again, we generalize them to a multivariate setting, impose equal-IFA constraints, and construct density forecasts.  

Using atmospheric CO$_2$ concentration data, the median of our preferred  first-NIFA distribution is 2031, with 80\% probability by 2033 and 95\% probability by 2034.  Using cumulative  CO$_2$ emissions data, the median is 2033, with 80\% probability by 2036 and 95\% probability by 2038. Therefore, it is very likely that the September Arctic  will be nearly ice-free at some time before the middle of the next decade.

\label{robust}Our  results are largely robust to  the modeling strategy (univariate or multivariate, carbon trend or direct time trend), the sea-ice indicator variable blended with $SIA$ ($SIE$, $SIT$, $SIV$), concentration path assumptions (high SSP5 8.5, medium SSP3 7.0, low SSP2 4.5), CO$_2$ measures (atmospheric concentration, cumulative emissions), and statistical confidence levels (80\%, 95\%).  Note in particular that although both this paper and \cite{DRice} agree that Arctic sea ice will disappear on a seasonal basis within a couple of decades, the two papers use very different information sets and methods, and the fact that they agree is itself an important result.  Hence, unless we somehow manage to follow something like the extremely low SSP1 2.6 concentration path, there is no escaping the sharp bottom line:  The Arctic will become seasonally ice free very soon -- most likely by  the mid- to late 2030's.

%
%***  Of course  only time will tell, but we think that our revised paper successfully makes the case that forecasts based on linear carbon trends, if certainly not irrefutable, are nevertheless worthy of very serious consideration.
%
In closing, we note that although this paper is about refining statistical sea-ice projections, not about  comparing statistical and climate-model  projections (unlike our earlier paper, \cite{DRice}), it nevertheless provides indirect  evidence supporting the speculations of, for example,  \cite{StroeveEtAl2007} and \cite{StroeveEtAl2012} that the slow decline in Arctic sea ice projected by climate models reflects an underestimation of the effects of GHGs. Our forecasts are indeed generally quite different from those of typical climate models; we predict an earlier first NIFA (in the 2030's), whereas the latest  CMIP6 coupled-climate models tend to obtain first NIFA around mid-century.  Hence, as regards future research, it will be of  interest to assess whether the   linear relationship between Arctic sea ice  and carbon dioxide, which we have documented in the observational record and used to make probabilistic assessments of NIFA arrival, is also present in simulated paths from large-scale dynamical climate models.  If so, it will be of great interest to assess whether the  ``$b$" parameters (as defined by our equation \eqref{equ:co2_uni}) embedded in various climate models are of sufficient magnitude to achieve consistency with the rapid observed historical Arctic sea-ice decline, and our projected continued rapid decline.   \cite{DRbivariate} take initial steps in that direction.

\appendix
\appendixpage
\addappheadtotoc
\newcounter{saveeqn}
\setcounter{saveeqn}{\value{section}}
\renewcommand{\theequation}{\mbox{\Alph{saveeqn}.\arabic{equation}}} \setcounter{saveeqn}{1}
\setcounter{equation}{0}

\section{Data sources and details} \label{data_app}

\renewcommand{\thefigure}{A\arabic{figure}}
\setcounter{figure}{0}

\renewcommand{\thetable}{A\arabic{table}}
\setcounter{table}{0}

\subsection{Measures of Arctic Sea Ice}

For the sample from 1979 to 2021, we consider four measures of September Arctic sea ice: area, extent, thickness, and volume.  

\subsubsection{Area and Extent}

Area data are from the National Snow and Ice Data Center (Sea Ice Index monthly dataset, Version 3, Dataset ID G02135, \url{https://nsidc.org/data/G02135/versions/3}). Area is from 30.98N, measured in $10^6\, {\rm km}^2$.  
The satellites miss the ``pole hole", and the published $SIA$ data exclude it, implicitly assuming that the pole hole has zero ice ($c{=}0$).  A better approximation is to assume that the pole hole has full ice ($c{=}1$).  Hence we first fill the pole hole by adding $1.19 \times 10^6 \, {\rm km}^2$ to $SIA$ from sample start through July 1987, $0.31 \times 10^6 \, {\rm km}^2$ from September 1987 through December 2007, and $0.029 \times 10^6 \, {\rm km}^2$ from January 2008 to present.\footnote{See the NSIDC Sea Ice Index Version 3 User Guide, \url{https://nsidc.org/data/G02135/versions/3}, p. 17.}

Extent data are from the National Snow and Ice Data Center (Sea Ice Index monthly dataset, Version 3, Dataset ID G02135, \url{https://nsidc.org/data/G02135/versions/3}).  Extent is from 30.98N,  measured in $10^6 \, {\rm km}^2$.  

\subsubsection{Thickness and Volume}

Thickness and volume date are from PIOMAS at the Polar Science Center, \url{http://psc.apl.uw.edu/research/projects/arctic-sea-ice-volume-anomaly/data/}.  Thickness is from 49N,   measured in m, where greater than 0.15 m.  Volume is from 49N,  measured in $10^3 \, {\rm km}^3$.  Volume data are published monthly. Thickness data are published on daily (file PIOMAS.thick.daily.1979.2020.Current.v2.1-2.dat), and we transform the daily data into monthly averages.

\subsection{CO$_2$ Measures}

The CO$_2$ measures are annual September atmospheric concentration and cumulative emissions.

\subsubsection{Concentration}

The historical (1979-2021) CO$_2$ atmospheric concentration data (in parts per million, PPM) are taken from NOAA Global Monitoring Laboratory \footnote{See: \url{https://gml.noaa.gov/webdata/ccgg/trends/co2/co2_mm_mlo.txt}. The data file was created on 11/05/2021 at 10:28:56. The data were retrieved on 11/23/2021. Our calculations are based on the \textit{de-seasonalized} data column.}  \textit{CO2C} is measured at Mauna Loa Observatory, Hawaii.

Scenario values for the period 2022-2100 are taken from the SSP Public Database (Version 2.0) (\url{https://tntcat.iiasa.ac.at/SspDb}). The individual scenarios are retrieved from the \textit{IAM Scenarios} spreadsheet as follows:
Data for the SSP1 2.6 scenario are based on the \textit{IMAGE - SSP1-26} model with variable ID \textit{Diagnostics|MAGICC6|Concentration|CO2} for region \textit{World}. 
Data for the SSP2 4.5 scenario are based on the \textit{MESSAGE-GLOBIOM - SSP2-45} model with variable ID \textit{Diagnostics|MAGICC6|Concentration|CO2} for region \textit{World}. 
Data for the SSP3 7.0 scenario are based on the \textit{AIM/CGE - SSP3-Baseline} model with variable ID \textit{Diagnostics|MAGICC6|Concentration|CO2} for region \textit{World}. 
Data for the SSP5 8.5 scenario are based on the \textit{REMIND-MAGPIE - SSP5-Baseline} model with variable ID \textit{Diagnostics|MAGICC6|Concentration|CO2} for region \textit{World}. 
All four scenarios are composed of eleven data points between 2005 and 2100. Missing data points for the years 2022-2100 were filled by linear interpolation.

\subsubsection{Emissions}

Cumulative CO$_2$ emissions data are taken from \cite{IPCC2021_data} and the corresponding IPCC report \citep{IPCC2021_policy}. The actual dataset \citep{IPCC2021_data} is at  \url{https://data.ceda.ac.uk/badc/ar6_wg1/data/spm/spm_10/v20210809}. From the latter, we use the following files:

\begin{enumerate}
	\item \textit{Top\_panel\_HISTORY.csv}, 
	\item \textit{Top\_panel\_SSP1-26.csv},
	\item \textit{Top\_panel\_SSP2-45.csv}, 
	\item \textit{Top\_panel\_SSP3-70.csv}, 
	\item \textit{Top\_panel\_SSP5-85.csv}.
\end{enumerate} 

\noindent We merge annual historical cumulative emissions for the period 1850-2019 with data from each of the four scenarios (SSP1 2.6, SSP2 4.5, SSP3 7.0, SSP5 8.5) from 2020 until 2050. 
Since the time-series end in 2050, we merge this data with the corresponding series from the SSP Public Database (Version 2.0) (\url{https://tntcat.iiasa.ac.at/SspDb}) to expand all four series across the second half of the century from 2051 to 2100. The individual scenarios are retrieved from the \textit{CMIP6 Emissions} spreadsheet as follows:
Data for the SSP1 2.6 scenario are based on the \textit{IMAGE - SSP1-26} model for region \textit{World}
Data for the SSP2 4.5 scenario are based on the \textit{MESSAGE-GLOBIOM - SSP2-45} model for region \textit{World}. 
Data for the SSP3 7.0 scenario are based on the \textit{AIM/CGE - SSP3-70 (Baseline)} model for region \textit{World}. 
Data for the SSP5 8.5 scenario are based on the \textit{REMIND-MAGPIE - SSP5-85 (Baseline)} model for region \textit{World}.
This leaves us with four annual time series 1850-2100, with a common historical part 1850-2019.

\section{Details of Bootstrap Simulation} \label{bootstrap}

Our bivariate bootstrap projection interval simulation  procedure, of which our univariate procedure is a special case, proceeds as follows at bootstrap replication $i$.  First we estimate the model and collect residuals. Second, we draw pairs of those residuals, to preserve their cross-correlation, and we add them back to the in-sample fitted values.  Third, we re-estimate the model on this synthetic data and use the re-estimated model to project out-of-sample. Finally, we draw pairs of residuals and add them to the forecasted paths to get the simulated path realizations. We repeat this for $i = 1, 2, ..., 1000$.  We then sort the simulated path values at each date and obtain the 90\% intervals shown in Figures \ref{fig:CO2_IceTimezzz} and  	\ref{fig:TIME}, with left and right endpoints given by the fifth and ninety-fifth percentiles, respectively, of the simulated path values.

Our bivariate bootstrap first NIFA simulation proceeds identically, with one extra step.  At the end, once the simulated path realizations are in hand, we calculate the corresponding simulated first  NIFA realizations, $\text{NIFA}_i$, $i = 1, 2, ..., 1000$.  We then obtain the  first NIFA densities shown in Figures \ref{fig:CO2_IceTimezzz} and \ref{fig:TIME} by applying a nonparametric kernel density estimator to $\text{NIFA}_i$, $i = 1, 2, ..., 1000$. 

\label{boot2} One might naturally wonder whether our bootstrap should account for dependence, which it does not.  Fortunately, however, the residuals in our annual models of September Arctic sea ice display negligible  dependence.  Indeed  $\rho(1)=0.06$ for both  linear carbon-trend residuals and quadratic time-trend residuals. (The key is to recognize that although for obvious reasons September shocks are likely to be highly correlated with those of earlier or later \textit{months} like August or October, they are much less likely to be highly correlated with those of earlier or later \textit{Septembers}.)

\bibliographystyle{Diebold}
\addcontentsline{toc}{section}{References}
\bibliography{Bibliography}

@article{BrockMiller2023,
  title={Polar Amplification in a Moist Energy Balance Model:
A Structural Econometric Approach to Estimation and Testing},
  author={Brock, W.A. and Miller, J.I.},
  note={Working Paper, Departments of Economics, University of Wisconson and University of Missouri},
  year={2023},
}

@article{eisenman2009nonlinear,
  title={Nonlinear Threshold Behavior During the Loss of Arctic Sea Ice},
  author={Eisenman, I. and Wettlaufer, J.S.},
  journal={Proceedings of the National Academy of Sciences},
  volume={106},
  pages={28--32},
  year={2009},
  publisher={National Academy of Sciences}
}

@article{tietsche2011recovery,
  title={Recovery Mechanisms of Arctic Summer Sea Ice},
  author={Tietsche, S. and Notz, D. and Jungclaus, J.H. and Marotzke, J.},
  journal={Geophysical Research Letters},
  volume={38},
  year={2011},
  publisher={Wiley Online Library}
}

@article{wang2023mechanisms,
  title={Mechanisms and Impacts of Earth System Tipping elements},
  author={Wang, S. and Foster, A. and Lenz, E.A. and Kessler, J.D. and Stroeve, J.C. and Anderson, L.O. and Turetsky, M. and Betts, R. and Zou, S. and Liu, W. and Boos, W. and Hausfather, Z.},
  journal={Reviews of Geophysics},
  volume={61},
  pages={e2021RG000757},
  year={2023},
  publisher={Wiley Online Library}
}

@article{winton2011climate,
  title={Do Climate Models 
mate the Sensitivity of Northern Hemisphere Sea Ice Cover?},
  author={Winton, M.},
  journal={Journal of Climate},
  volume={24},
  pages={3924--3934},
  year={2011},
  publisher={American Meteorological Society}
}

@article{raupach2013exponential,
  title={The Exponential Eigenmodes of the Carbon-Climate System, and their Implications for Ratios of Responses to Forcings},
  author={Raupach, M.R.},
  journal={Earth System Dynamics},
  volume={4},
  pages={31-49},
  year={2013},
  publisher={Copernicus GmbH}
}

@article{bennedsen2023multivariate,
  title={A Multivariate Dynamic Statistical Model of the Global Carbon Budget 1959-2020},
  author={Bennedsen, M. and Hillebrand, E. and Koopman, S.J.},
  journal={Journal of the Royal Statistical Society Series A: Statistics in Society},
  volume={186},
  pages={20-42},
  year={2023},
  publisher={Oxford University Press US}
}

@inproceedings{bacastow1973changes,
  author = {Bacastow, R and Keeling, C.D.},
  title = {Atmospheric Carbon Dioxide and Radiocarbon in the Natural Cycle: II. Changes from 1700 to 2070 as Deduced from a Geochemical Model},
  note = {in \emph {Carbon and the Biosphere Conference Proceedings. Brookhaven Symposia in Biology}, Upton, New York, 86--135},
  year = {1973},
%  pages = {86--135},
  %address = {Upton, New York },
}

@article{budyko1969effect,
  title={The Effect of Solar Radiation Variations on the Climate of the Earth},
  author={Budyko, M.I.},
  journal={Tellus},
  volume={21},
  pages={611--619},
  year={1969},
  publisher={Taylor \& Francis}
}

@article{sellers1969global,
  title={A Global Climatic Model based on the Energy Balance of the Earth-Atmosphere System},
  author={Sellers, W.D.},
  journal={Journal of Applied Meteorology and Climatology},
  volume={8},
  pages={392--400},
  year={1969}
}

@article{gregory2000vertical,
  title={Vertical Heat Transports in the Ocean and their Effect on Time-Dependent Climate Change},
  author={Gregory, J.M.},
  journal={Climate Dynamics},
  volume={16},
  pages={501--515},
  year={2000},
  publisher={Springer}
}

@article{held2010probing,
  title={Probing the Fast and Slow Components of Global Warming by Returning Abruptly to Preindustrial Forcing},
  author={Held, I.M. and Winton, M. and Takahashi, K. and Delworth, T. and Zeng, F. and Vallis, G.K.},
  journal={Journal of Climate},
  volume={23},
  pages={2418--2427},
  year={2010},
  publisher={American Meteorological Society}
}

@article{hillebrand2023,
  title={Discussion of `When Will Arctic Sea Ice Disappear? Projections of Area, Extent, Thickness and Volume', by F.X. Diebold, G.D. Rudebusch, M. Goebel, P. Goulet Coulombe, and B. Zhang},
  author={Hillebrand, E.},
  year={2023},
  note={Conference on Climate Risk, Society for Financial Econometrics, April}
}

@article{StroeveEtAl2012,
        Author = {Stroeve, J.C. and Serreze, M.C. and Holland, M.M. and Kay, J.E. and Malanik, J. and Barrett, A.P.},
        Doi = {10.1007/s10584-011-0101-1},
        Journal = {Climatic Change},
        Number = {3},
        Pages = {1005--1027},
        Title = {The Arctic's Rapidly Shrinking Sea Ice Cover: A Research Synthesis},
        Volume = {110},
        Year = {2012}
}

@article{StroeveEtAl2007,
author = {Stroeve, J. and Holland, M.M. and Meier, W. and Scambos, T. and Serreze, M.},
title = {Arctic Sea Ice Decline: Faster than Forecast},
journal = {Geophysical Research Letters},
volume = {34},
number = {9},
pages = {},
doi = {https://doi.org/10.1029/2007GL029703},
year = {2007}
}

@article{Stock1994,
    author = {Stock, J.H.},
    title = "{Chapter 46 Unit Roots, Structural Breaks and Trends}",
    journal = {Handbook of Econometrics, Elsevier},
    year = {1994},
    volume = {4},
    doi = {https://doi.org/10.1016/S1573-4412(05)80015-7}
}

@article{NelsonPlosser1982,
title = "{Trends and Random Walks in Macroeconmic Time Series: Some Evidence and Implications}",
journal = {Journal of Monetary Economics},
volume = {10},
number = {2},
pages = {139-162},
year = {1982},
doi = {https://doi.org/10.1016/0304-3932(82)90012-5},
author = {Nelson, C.R. and Plosser, C.R.}
}

@article{Perron1989,
 author = {Perron, P.},
 journal = {Econometrica},
 number = {6},
 pages = {1361--1401},
 title = {The Great Crash, the Oil Price Shock, and the Unit Root Hypothesis},
 volume = {57},
 year = {1989}
}

@article{NotzSIMIP2020,
author = {Notz, D. and SIMIP, Community},
title = {Arctic Sea Ice in CMIP6},
journal = {Geophysical Research Letters},
volume = {47},
number = {10},
doi = {https://doi.org/10.1029/2019GL086749},
url = {https://agupubs.onlinelibrary.wiley.com/doi/abs/10.1029/2019GL086749},
year = {2020}
}

@article{LynchNorchiLi2022,
author = {Lynch, A.H. and Norchi, C.H. and Li, X.},
title = {The Interaction of Ice and Law in Arctic Marine Accessibility},
journal = {Proceedings of the National Academy of Sciences},
volume = {119},
number = {26},
pages = {e2202720119},
year = {2022},
doi = {10.1073/pnas.2202720119},
URL = {https://www.pnas.org/doi/abs/10.1073/pnas.2202720119}
}

@article{IPCC2021_Ch9,
	author = {Fox-Kemper, B. and Hewitt, H.T. and Xiao, C. and Adalgeirsdottir, G. and Drijfhout, S.S. and Edwards, T.L. and Golledge, N.R. and Hemer, M. and Kopp, R.E. and Krinner, G. and Mix, A. and Notz, D. and Nowicki, S. and Nurhati, I.S. and Ruiz, L. and Sallée, J-B. and Slangen, A.B.A. and Yu, Y.},
	title = {Ocean, Cryosphere and Sea Level Change},
	note = {in Masson-Delmotte, V., P. Zhai, A. Pirani, S.L. Connors, C. Péan, S. Berger, N. Caud, Y. Chen, L. Goldfarb, M.I. Gomis, M. Huang, K. Leitzell, E. Lonnoy, J.B.R. Matthews, T.K. Maycock, T. Waterfield, O. Yelekçi, R. Yu, and B. Zhou, eds., \textit{Climate Change 2021: The Physical Science Basis. Contribution of Working Group I to the Sixth Assessment Report of the Intergovernmental Panel on Climate Change}, Cambridge University Press, in press. \url{https://ipcc.ch/report/ar6/wg1/downloads/report/IPCC_AR6_WGI_SPM_final.pdf}},
	year = {2022}
}

@article{Alvarez2020,
	Author = {Alvarez, J. and Yumashev, D. and Whiteman, G.},
	Doi = {10.1007/s13280-019-01211-z},
	Journal = {Ambio},
	Number = {2},
	Pages = {407--418},
	Title = {A Framework for Assessing the Economic Impacts of Arctic Change},
	Url = {https://doi.org/10.1007/s13280-019-01211-z},
	Volume = {49},
	Year = {2020}
}

@article{IPCC2021_data,
  author = {Rogelj, J. and Trewin, B. and Haustein, K. and Canadell, P. and Szopa, S. and Milinski, S. and Marotzke, J. and Zickfeld, K.},
  title = {Summary for Policymakers of the Working Group I Contribution to the IPCC Sixth Assessment Report - Data for Figure SPM.10 (v20210809)},
note={NERC EDS Centre for Environmental Data Analysis},
  url={http://dx.doi.org/10.5285/cfe938e70f8f4e98b0622296743f7913},
  year = {2021}
}

@article{IPCC2021_policy,
	author = {Allan, R.P. and Hawkins, E. and Ballouin, N. and Collins, B.},
	title = {IPCC, 2021: Summary for Policymakers},
	note = {in Masson-Delmotte, V., P. Zhai, A. Pirani, S.L. Connors, C. Péan, S. Berger, N. Caud, Y. Chen, L. Goldfarb, M.I. Gomis, M. Huang, K. Leitzell, E. Lonnoy, J.B.R. Matthews, T.K. Maycock, T. Waterfield, O. Yelekçi, R. Yu, and B. Zhou, eds., \textit{Climate Change 2021: The Physical Science Basis. Contribution of Working Group I to the Sixth Assessment Report of the Intergovernmental Panel on Climate Change}, Cambridge University Press, in press. \url{https://ipcc.ch/report/ar6/wg1/downloads/report/IPCC_AR6_WGI_SPM_final.pdf}},
	year = {2022}
}

@article{Mudryk2021,
	author = {Mudryk, L.R. and Dawson, J. and Howell, S.E.L. and Derksen, C. and Zagon, T.A. and Brady, M.},
	title = {Impact of 1, 2 and 4$^{\circ}$C of Global Warming on Ship Navigation in the Canadian Arctic},
	journal = {Nature Climate Change},
  volume = {11},
  number = {8},
  pages = {673--679},
  year = {2021},
  publisher = {Nature Publishing Group}
}

@article{IPCC2021tech,
	author = {Arias, P.A. et al.},
	title = {Technical Summary},
	note = {in Masson-Delmotte, V., P. Zhai, A. Pirani, S.L. Connors, C. Péan, S. Berger, N. Caud, Y. Chen, L. Goldfarb, M.I. Gomis, M. Huang, K. Leitzell, E. Lonnoy, J.B.R. Matthews, T.K. Maycock, T. Waterfield, O. Yelekçi, R. Yu, and B. Zhou, eds., \textit{Climate Change 2021: The Physical Science Basis. Contribution of Working Group I to the Sixth Assessment Report of the Intergovernmental Panel on Climate Change}, Cambridge University Press, in press. \url{https://www.ipcc.ch/report/ar6/wg1/downloads/report/IPCC_AR6_WGI_TS.pdf}},
	year = {2022}
}

@article{Zygmuntowskaetal2014,
	author = {Zygmuntowska, M. and Rampal, P. and Ivanova, N. and Smedsrud, L.H.},
	title = {Uncertainties in Arctic Sea Ice Thickness and Volume: New Estimates and Implications for Trends},
	journal = {The Cryosphere},
	volume={8},
	number={},
	pages={705-720},
	year = {2014}
}

@article{Schweigeretal2011,
author = {Schweiger, A. and Lindsay, R. and Zhang, J. and Steele, M. and Stern, H. and Kwok, R.},
title = {Uncertainty in Modeled Arctic Sea Ice Volume},
journal = {Journal of Geophysical Research: Oceans},
volume = {116},
number = {C8},
pages = {C00D06},
doi = {10.1029/2011JC007084},
url = {https://agupubs.onlinelibrary.wiley.com/doi/abs/10.1029/2011JC007084},
eprint = {https://agupubs.onlinelibrary.wiley.com/doi/pdf/10.1029/2011JC007084},
year = {2011}
}

@Article{Selyuzhenoketal2020,
AUTHOR = {Selyuzhenok, V. and Bashmachnikov, I. and Ricker, R. and Vesman, A. and Bobylev, L.},
TITLE = {Sea Ice Volume Variability and Water Temperature in the Greenland Sea},
JOURNAL = {The Cryosphere},
VOLUME = {14},
YEAR = {2020},
NUMBER = {2},
PAGES = {477-495},
URL = {https://tc.copernicus.org/articles/14/477/2020/},
DOI = {10.5194/tc-14-477-2020}
}

@article{Chevallieretal2017,
	author = {Chevallier, M. and Smith, G.C. and Dupont, F. and Lemieux, J.F. and Forget, G. and Fujii, Y. and Hernandez, F. and Msadek, R. and Peterson, K.A. and Storto, A. and Toyoda, T.},
	title = {Intercomparison of the Arctic Sea Ice Cover in Global Ocean-Sea Ice Reanalyses From the ORA-IP Project},
	journal = {Climate Dynamics},
	volume={49},
	number={3},
	pages={1107-1136},
	year = {2017}
}

@article{Bunzeletal2018,
	author = {Bunzel, F. and Notz, D. and Pedersen, L.T.},
	title = {Retrievals of Arctic Sea-Ice Volume and its Trend Significantly Affected by Interannual Snow Variability},
	journal = {Geophysical Research Letters},
	volume={45},
	number={21},
	pages={11751-11759},
	year = {2018}
}

@article{Cooley2020,
  title={Coldest Canadian Arctic Communities Face Greatest Reductions in Shorefast Sea Ice},
  author = {Cooley, S.W. and Ryan, J.C. and Smith, L.C. and Horvat, C. and Pearson, B. and Dale, B. and Lynch, A.H.},
  journal={Nature Climate Change},
  volume={10},
  number={6},
  pages={533--538},
  year={2020},
  publisher={Nature Publishing Group}
}

@article{SeaIceBookCh8,
	title={Sea Ice Modelling},
	author={Lepp\"{a}ranta, M. and Meleshko, V.P. and Uotila, P. and Pavlova, T.},
	note={in Johannessen, O.M., Bohylev, L.P., Shalina, E.V., and Sandven, S., eds., \emph{Sea Ice in the Arctic: Past Present and Future}, 318-388, Springer Nature},
	year={2020},
}

@article{SeaIceBookCh5,
	title={Arctic Sea Ice Thickness and Volume Transformation },
	author={Shalina, E.V. and Khvorostovsky, K. and Sandven, S.},
	note={in Johannessen, O.M., Bohylev, L.P., Shalina, E.V., and Sandven, S., eds., \emph{Sea Ice in the Arctic: Past Present and Future}, 167-246, Springer Nature},
	year={2020},
}

@article{Rosenblum2017,
	author = {Rosenblum, E. and Eisenman, I.},
	title = {Sea Ice Trends in Climate Models Only Accurate in Runs with 49 Biased Global Warming},
	journal = {Journal of Climate},
	volume = {30},
	number = {16},
	pages = {6265-6278},
	year = {2017},
}

@article{Jahn2016,
author = {Jahn, A. and Kay, J.E. and Holland, M.M. and Hall, D.M.},
title = {How Predictable is the Timing of a Summer Ice-Free Arctic?},
journal = {Geophysical Research Letters},
volume = {43},
number = {17},
pages = {9113-9120},
year = {2016},
}

@article{WangOverland2009,
author = {Wang, M. and Overland, J.E.},
title = {A Sea Ice Free Summer Arctic Within 30 Years?},
journal = {Geophysical Research Letters},
volume = {36},
number = {7},
pages = {L07502},
year = {2009},
}

@article{ZhangRoth03,
author = {Zhang, J. and Rothrock, D.A.},
title = {Modeling Global Sea Ice with a Thickness and Enthalpy Distribution Model in Generalized Curvilinear Coordinates},
journal = {Monthly Weather Review},
volume = {131},
number = {5},
pages = {845-861},
year = {2003},
doi = {10.1175/1520-0493(2003)131<0845:MGSIWA>2.0.CO;2},
}

@article{Wangetal2016,
author = {Wang, Lei and Yuan, Xiaojun and Ting, Mingfang and Li, Cuihua},
title = {Predicting Summer Arctic Sea Ice Concentration Intraseasonal Variability Using a Vector Autoregressive Model},
journal = {Journal of Climate},
volume = {29},
number = {4},
pages = {1529-1543},
year = {2016},
doi = {10.1175/JCLI-D-15-0313.1},
URL = {https://doi.org/10.1175/JCLI-D-15-0313.1}
}

@article{SeaIceBookCh4,
	title={Changes in Arctic Sea Ice Cover in the Twentieth and Twenty-First Centuries},
	author={Shalina, E.V. and Johannessen, O.M.  and Sandven, S.},
	note={in Johannessen, O.M., Bohylev, L.P., Shalina, E.V., and Sandven, S., eds., \emph{Sea Ice in the Arctic: Past Present and Future}, 93-166, Springer Nature},
	year={2020},
}

@article {SSW,
	title = {Inference in Linear Time Series Models with Some Unit Roots},
	journal = {Econometrica},
	volume = {58},
	year = {1990},
	pages = {113-144},
	author = {Sims, C.A. and Stock, J.H. and Watson, M.W.}
}

@article{DRice,
  title={Probability Assessments of an Ice-Free Arctic: Comparing Statistical and Climate Model Projections},
  author={Diebold, F.X. and Rudebusch, G.D.},
  year={2022},
journal={Journal of Econometrics},
volume={231},
pages={520-524}
}

@article{schweiger2021,
	title={Accelerated Sea Ice Loss in the Wandel Sea Points to a Change in the Arctic’s Last Ice Area},
	author={Schweiger, A.J. and S. Michael and J. Zhang and G.W.K. Moore and K.L. Laidre},
	journal={Communications Earth \& Environment},
	volume={2},
	number={1},
	pages={1--11},
	year={2021},
	publisher={Nature Publishing Group}
}

@article{Johannessen2008,
	author = {Johannessen, O.M.},
	title = {Decreasing Arctic Sea Ice Mirrors Increasing CO2 on Decadal Time Scale},
	journal = {Atmospheric and Oceanic Science Letters},
	volume = {1},
	number = {1},
	pages = {51-56},
	year = {2008},
}

@article{StroeveNotz2018,
	author = {Stroeve, J. and D. Notz},
	title = {Changing State of Arctic Sea Ice Across All Seasons},
	journal = {Environmental Research Letters},
	volume = {13},
	number = {10},
	pages = {103001},
	year = {2018}
}

@article{NS2016,
	author = {Notz, D. and Stroeve, J.},
	title = {Observed Arctic Sea-Ice Loss Directly Follows Anthropogenic CO2 Emission},
	journal = {Science},
	volume = {354},
	number = {6313},
	pages = {747-750},
	year = {2016},
}

@article{DRbivariate,
	author={Diebold, F.X. and Rudebusch, G.D.},
	title   = {Climate Models Underestimate the Sensitivity of Arctic Sea Ice to Carbon Emissions},
	note    = {Manuscript, University of Pennsylvania},
	year    = {2023},
}

@article{Labeetal2018,
	title={Variability of Arctic Sea Ice Thickness Using PIOMAS and the CESM Large Ensemble},
	author={Labe, Z. and Magnusdottir, G. and Stern, H.},
	journal={Journal of Climate},
	volume={31},
	number={8},
	pages={3233--3247},
	year={2018},
	publisher={American Meteorological Society}
}

@article{Chylek2022,
	title={Annual Mean Arctic Amplification 1970-2020: Observed and Simulated by CMIP6 Climate Models},
	author={Chylek, P. and Folland, C. and Klett, J.D. and Wang, M. and Hengartner, N. and Lesins, G. and Dubey, M.K.},
	journal={Geophysical Research Letters},
	volume = {49},
	number = {13},
	pages={e2022GL099371},
	year={2022},
	publisher={Wiley Online Library}
}

@article{Kaufmanetal2006a,
	title={Emissions, Concentrations, \& Temperature: A Time Series Analysis},
	author={Kaufmann, R.K. and Kauppi, H. and Stock, J.H.},
	journal={Climatic Change},
	volume={77},
	number={3},
	pages={249--278},
	year={2006},
	publisher={Springer}
}

@article{PerronEstrada2019,
	title={Breaks, Trends and the Attribution of Climate Change: A Time-Series Analysis},
	author={Estrada, F. and Perron, P.},
	journal={Economia},
	volume={42},
	number={83},
	pages={1--31},
	year={2019}
}

@article{MillerNam2020,
	title={Dating Hiatuses: A Statistical Model of the Recent Slowdown in Global Warming and the Next One},
	author={Miller, J.I. and Nam, K.},
	journal={Earth System Dynamics},
	volume={11},
	number={4},
	pages={1123--1132},
	year={2020},
	publisher={Copernicus GmbH}
}

@article{Vaksetal2020,
	title={Palaeoclimate Evidence of Vulnerable Permafrost during Times of Low Sea Ice},
	author={Vaks, A. and Mason, A.J. and Breitenbach, S.F.M. and Kononov, A.M. and Osinzev, A.V. and Rosensaft, M. and Borshevsky, A. and Gutareva, O.S. and Henderson, G.M.},
	journal={Nature},
	volume={577},
	number={7789},
	pages={221--225},
	year={2020},
	publisher={Nature Publishing Group}
}

\end{document}